\documentclass[]{article}

\usepackage{widetext}
\usepackage{graphicx}
\usepackage{caption}
\usepackage{subcaption}
\usepackage{lineno}

\usepackage{verbatim}
\usepackage{amsmath}
\usepackage{amssymb}
\usepackage{color}
\usepackage{hyperref}
 \usepackage{stackrel}
\usepackage{upgreek}
 \usepackage{csquotes}
 
 \usepackage{fullpage}

 \def\##1{{\bf #1}}
\renewcommand\vec{\mathbf}
\def\doubleunderline#1{\underline{\underline{#1}}}
\def\=#1{\underline{\underline{#1}}}

 \def\eps{\varepsilon}

 \def\epso{\eps_{\scriptscriptstyle 0}}

\def\muo{\mu_{\scriptscriptstyle 0}}
\def\ko{k_{\scriptscriptstyle 0}}
\def\koc{{\ko}c}

\def\etao{\eta_{\scriptscriptstyle 0}}
\def\epsr{\eps_{r}}
\def\mur{\mu_{r}}
\def\etar{\eta_{r}}
\def\ux{\hat{\#x}}
\def\uy{\hat{\#y}}
\def\uz{\hat{\#z}}
\def\ur{\hat{\#r}}
\def\uphi{\hat{ \boldsymbol\phi}}
\def\utheta{\hat{ \boldsymbol\theta}}
\def\e{\hat{\#e}}

\def\k{\hat{\#k}}
\def\nhat{\hat{\#n}}

\def\ep{\hat{\#e}^{\parallel}}
\def\es{\hat{\#e}^{\perp}}
\def\eLCP{\hat{\#e}^{\rm LCP}}
\def\eRCP{\hat{\#e}^{\rm RCP}}

\def\.{\mbox{ \tiny{$^\bullet$} }}

\newenvironment{rcases}
  {\left.\begin{aligned}}
  {\end{aligned}\right\rbrace}
  
  \def\les{\left[}
\def\ris{\right]}
\def\lec{\left\{}
\def\ric{\right\}}
  
\def\GerM{\mbox{\boldmath${\mathfrak M}$}}
\def\GerN{\mbox{\boldmath${\mathfrak N}$}}

\def\newJ{{\mathfrak J}}
\def\newK{{\mathfrak K}}

\def\calU{{\cal U}}
\def\calV{{\cal V}}  
\def\calS{{\cal S}}

\def\dprime{{\prime \prime}}

\def\one{^{(1)}}
\def\two{^{(2)}}

\def\inc{_{\rm inc}}

\def\smn{_{smn}}
\def\oneparl{^{(1)\parallel}}
\def\oneperp{^{(1)\perp}}
\def\oneLCP{^{(1){\rm L}}}
\def\oneRCP{^{(1){\rm R}}}
\def\threeparl{^{(3)\parallel}}
\def\threeperp{^{(3)\perp}}
\def\threeLCP{^{(3){\rm L}}}
\def\threeRCP{^{(3){\rm R}}}

\def\QD{Q_D}
\def\Qsca{Q_{\text{sca}}}
\def\Qabs{Q_{\text{abs}}}
\def\Qext{Q_{\text{ext}}}
\def\Qb{Q_b}

\def\Qscaparl{Q_{\rm sca}^{\parallel}}
\def\Qscaperp{Q_{\rm sca}^{\perp}}
\def\QscaLCP{Q_{\rm sca}^{\rm L}}
\def\QscaRCP{Q_{\rm sca}^{\rm R}}
\def\Qextparl{Q_{\rm ext}^{\parallel}}
\def\Qextperp{Q_{\rm ext}^{\perp}}
\def\QextLCP{Q_{\rm ext}^{\rm L}}
\def\QextRCP{Q_{\rm ext}^{\rm R}}
\def\Qabsparl{Q_{\rm abs}^{\parallel}}
\def\Qabsperp{Q_{\rm abs}^{\perp}}
\def\QabsLCP{Q_{\rm abs}^{\rm L}}
\def\QabsRCP{Q_{\rm abs}^{\rm R}}

\begin{document}

\begin{center}
\textbf{Plane-wave scattering by an ellipsoid composed of an orthorhombic
dielectric-magnetic material with arbitrarily oriented constitutive principal axes}\\
\textit{Hamad M. Alkhoori}\\
{Department of Electrical Engineering, The Pennsylvania State University, University Park, Pennsylvania 16802, USA}\\
\textit{Akhlesh Lakhtakia}\\
{Department of Engineering Science and Mechanics, The Pennsylvania State University, University Park, Pennsylvania 16802, USA}\\
\textit{James K. Breakall}\\
{Department of Electrical Engineering, The Pennsylvania State University, University Park, Pennsylvania 16802, USA}\\
\textit{Craig F. Bohren}\\
{Department of Meteorology, The Pennsylvania State University, University Park, Pennsylvania 16802, USA} 
 \end{center}

\begin{abstract}
The extended boundary condition method can be formulated to study plane-wave scattering by an ellipsoid composed of
an orthorhombic dielectric-magnetic material whose relative permittivity dyadic is a scalar multiple of its relative
permeability dyadic, when  the constitutive principal axes are arbitrarily oriented with respect to the shape
principal axes. Known vector spherical wavefunctions are used to represent the fields in the surrounding matter-free space. After deriving closed-form expressions for the vector spherical wavefunctions for the scattering material,
the internal fields are represented as  superpositions of those vector spherical wavefunctions.
The unknown scattered-field coefficients are related to the known incident-field coefficients by a transition matrix. The total  scattering   and   absorption efficiencies  are highly affected by the orientation of the constitutive principal axes relative to the shape principal axes, and the effect of the orientational mismatch between the two sets of principal axes  is more pronounced as the electrical size increases.
  The dependence of the
total scattering efficiency, but not of the absorption efficiency, on the angle of rotation about a shape principal
axis can be predicted qualitatively from the variation of a scalar function with respect to the angle of rotation. The  total scattering and absorption efficiencies do not depend on the polarization state of incident plane wave when the scattering material is impedance-matched to free space. The polarization state of the incident plane wave has a more discernible effect on the total scattering   and absorption efficiencies for ellipsoids compared to spheres.  
\end{abstract}
  
\section{Introduction}
Analysis of frequency-domain electromagnetic fields in a homogeneous dielectric-magnetic material require the specification of two constitutive dyadics, one of which relates $\vec{D}(\vec{r})$ to $\vec{E}(\vec{r})$
and the other relates $\vec{B}(\vec{r})$ to $\vec{H}(\vec{r})$.   Each constitutive dyadic comprises nine complex  components in general \cite{Post,Lakhtakiaan}. It is convenient to choose a coordinate system in which
at least one of the two constitutive dyadics is diagonal \cite{Chen,Berry}. Any symmetric dyadic can be diagonalized, and diagonalization can be effected through a rotation of axes in some situations 
\cite[][p.~202]{collin}. Rotation transforms a problem from a laboratory coordinate system into a coordinate system in which a constitutive dyadic is diagonal. The eigenvalues of a diagonal dyadic populate its diagonal and the corresponding eigenvectors determine the principal axes of that dyadic.
 
Obtaining closed-form expressions for the electromagnetic field
phasors in a homogeneous dielectric-magnetic material is generally difficult,
all the more so if Cartesian coordinates are not used \cite{Chen,FLbook}.
Recently, however, closed-form expressions of vector spherical wavefunctions were obtained for an orthorhombic dielectric-magnetic material whose relative permittivity dyadic
 is a scalar multiple of its relative permeability
dyadic \cite{Lakhtakiaoriginal}.  These  wavefunctions were used to formulate a scattering problem wherein an incident field impinges on an object composed of the chosen material and suspended
in free space (i.e., vacuum). The formulation is based on  the extended boundary condition method (EBCM), also called the null-field method and the T-matrix method \cite{Waterman, Lakh-IskBook}. This method requires knowledge of (i) the bilinear expansions of the dyadic Green functions for free
space \cite{Morse,BY1975} and (ii) the vector wavefunctions to completely express the field phasors induced inside the object.

The frequency-domain constitutive relations of the chosen material are \cite{Lakhtakiaoriginal}
\begin{equation} \label{D}
 \vec{D}(\vec{r})=\epso \epsr \={C}  \. \vec{E}(\vec{r})
 \end{equation}
 and
 \begin{equation} \label{B}
 \vec{B}(\vec{r})=\muo \mur \={C}  \.  \vec{H}(\vec{r})\,,
\end{equation}
where $\epso$ and $\muo$ are the permittivity and permeability of free space, respectively;  $\epsr$ and $\mur$ are complex functions of the angular frequency $\omega$; and the dyadic
\begin{equation} \label{B_dyadic}
 \={C}= \={S}   \. \={A}  \. \={A} \. \={S}^{-1}\,.
\end{equation}
Here, 
\begin{equation}
\={S}= \={R}_z(\gamma) \. \={R}_y(\beta) \. \={R}_z(\alpha)
\label{S-def}
\end{equation}
is a product of three rotation dyadics, with
\begin{eqnarray} 
\nonumber
&&\={R}_z(\zeta)= (\ux \ux+ \uy \uy) \cos \zeta - (\ux \uy-\uy \ux) \sin \zeta + \uz \uz\,,
\\[5pt]
\label{transformation1_def}
&&\qquad\quad
\zeta\in\left\{\alpha,\gamma\right\}\,,
\end{eqnarray}
and
\begin{eqnarray} 
 \label{transformation2_def}
&&\={R}_y(\beta)= (\ux \ux+ \uz \uz) \cos \beta - (\uz \ux-\ux \uz) \sin \beta + \uy \uy\,.
\end{eqnarray}
The dyadic $\={R}_z(\alpha)$ represents a rotation by $\alpha\in [0,\pi]$ around the $z$ axis,  $\={R}_y(\beta)$ represents  a rotation by $\beta\in [0,\pi]$ around the $y$ axis, and  $\={R}_z(\gamma)$ represents a rotation by $\gamma\in [0,\pi]$ around the $z$ axis. 

The diagonal dyadic $\={A}$ is defined as 
\begin{equation} \label{mata}
\={A}= \alpha_x^{-1}\ux \ux + \alpha_y^{-1} \uy \uy + \uz \uz\,.
\end{equation}
The constitutive-anisotropy
parameters $\alpha_x$ and $\alpha_y$ are real positive functions of $\omega$ to ensure that $\={A}$ is positive definite \cite{Lutkepohl}, which allows for an affine transformation of space wherein fields can be compactly expressed \cite{Lakhtakiaoriginal}. By virtue of  Eqs.~(\ref{D})--(\ref{mata}), 
the permittivity dyadic 
$\={\eps}= \epso \epsr   \={C}$
and the permeability dyadic
$\={\mu}=  \muo \mur  \={C}$
have the same set of three principal axes.

In addition to the constitutive principal axes, an object has a shape. The ellipsoidal shape is convex and
 possesses three principal axes. In a coordinate system with its origin at the centroid of an ellipsoid,
the surface $\calS$ of that ellipsoid is delineated by the position vector
\begin{eqnarray}  
&&
\nonumber
\#r_s(\theta,\phi)=c\=U\. \left[\left(\ux\cos\phi+\uy\sin\phi\right)\sin\theta +\uz\cos\theta\right]\,,
\\[5pt]
&&\quad
\quad \theta\in[0,\pi]\,,\quad\phi\in[0,2\pi)\,.
\label{rs-def}
\end{eqnarray}
The  {shape dyadic}
\begin{equation}
\label{U-def}
\=U=\left(a\ux\ux+b\uy\uy\right)/c+\uz\uz\,
\end{equation}
contains the $x$, $y$, and $z$ axes as
the shape principal axes of an ellipsoid with linear dimensions  $2a, 2b$, and $2c$ along those axes.
We take the  laboratory coordinate system to be the one in which the shape dyadic $\=U$ is defined
via Eq.~(\ref{U-def}).

Whether or not materials described by Eqs.~(\ref{D})--(\ref{mata}) exist in nature,   
an effectively homogeneous material described by those equations  can 
potentially be fabricated  by properly dispersing
electrically small wires, loops, and other inclusions   of
different  materials and shapes in some host material \cite{WC,MW2000,MMR,Dionne}. Also,
 certain spacetime metrics can yield the chosen forms of $\=\eps$ and $\=\mu$
\cite{Plebanski}.  

In this paper, we consider scattering by an ellipsoid endowed with the shape dyadic $\=U$,
the permittivity dyadic $\=\eps$ and the permeability dyadic $\=\mu$. The scattering characteristics must depend on the orientation of the constitutive principal axes determined from $\={C}$ relative to the orientation of shape principal axes determined from  $\=U$. {Scattering by an ellipsoid composed of an orthorhombic dielectric-magnetic material has been previously studied \cite{Alkhoori1} for $\=S=\=I$,
but that constraint is relaxed in this paper.  In order to distinguish the effects of the orientation of the shape principal axes  from
those of the  orientation of the constitutive principal axes, comparison with a sphere composed of the same material \cite{Jafri,Jafri-err} is necessary.}

Solution of the boundary-value problem of scattering requires  closed-form expressions of the vector spherical wavefunctions in the laboratory coordinate system, but these are  available
 \cite{Lakhtakiaoriginal} only for $\=S=\=I$,
the identity dyadic \cite{Chen}. Therefore, we derived the expressions for $\=S\ne\=I$. In order to quantify the effects of orienting the constitutive principal axes differently from the shape principal axes, we computed the total scattering efficiency when the ellipsoid is illuminated by a plane wave \cite{Bohren,Alkhoori1}.

This paper is organized as follows.  
The  closed-form expressions of the vector spherical wavefunctions in  the laboratory coordinate system are presented in Sec.~\ref{s1} for the material specified by Eqs.~(\ref{D})--(\ref{mata}). In the same section, we also  solve the scattering problem using the EBCM. In Section \ref{s2}, we present and analyze computed values of the total scattering efficiency in relation to the orientations of the constitutive principal axes and shape principal axes. Our conclusions are summarized in Section \ref{s3}. 

An $\exp(-i \omega t)$ dependence on time $t$ is implicit  throughout the analysis with $i=\sqrt{-1}$. Vectors are  in boldface, unit vectors are decorated by caret,  dyadics are  double underlined, and column vectors as well as matrices are enclosed in square brackets. The free-space wavenumber is denoted by $\ko= \omega\sqrt{\muo \epso}$ and the intrinsic impedance of free space by $\etao=\sqrt{\muo / \epso}$.
We define   the relative impedance  $\etar =\sqrt{\mur/\epsr}$ and the wavenumber $k=\ko\sqrt{\epsr}\sqrt{\mur}\alpha_x^{-1}\alpha_y^{-1}$.

\section{Theory}\label{s1}
\subsection{Closed-form expressions of vector spherical wavefunctions} \label{vector_wavefunctions}
On substituting Eqs. (\ref{D}) and (\ref{B}) into the source-free Maxwell curl equations
$\nabla \times \vec{E}(\vec{r}) =  i \omega \vec{B}(\vec{r})$  and 
$\nabla \times \vec{H}(\vec{r}) =   -i \omega  \vec{D}(\vec{r})$, one obtains
\begin{equation} \label{Curl_eqs}
\begin{rcases}
\begin{aligned}
\nabla \times \vec{E}(\vec{r})& =  i \omega \big[ \muo \mur \={S} \. \={A} \. \={A} \. \={S}^{-1} \. \vec{H}(\vec{r}) \big]  \\ 
\nabla \times \vec{H}(\vec{r})& =   -i \omega  \big[ \epso \epsr \={S} \. \={A} \. \={A} \. \={S}^{-1} \. \vec{E}(\vec{r})  \big]   \\ 
\end{aligned}
\end{rcases}.
\end{equation} 
The complete solutions $\vec{E}(\vec{r})$ and $\vec{H}(\vec{r})$ of Eqs.~(\ref{Curl_eqs}) can be written as  superposition of vector spherical wavefunctions
\begin{equation} \label{closed-form_m}
 \GerM_{smn}^{(j)}(\vec{r}) = \={S} \. \={A}^{-1} \. \vec{M}_{smn}^{(j)}(k \={A}^{-1} \.  \={S} \. \vec{r}) 
 \end{equation}
 and
 \begin{equation} \label{closed-form_n}
  \GerN_{smn}^{(j)}(\vec{r}) = \={S} \. \={A}^{-1}\. \vec{N}_{smn}^{(j)}(k \={A}^{-1} \.  \={S} \. \vec{r})\,,
\end{equation}
as proved elsewhere~\cite{Lakhtakiaoriginal,Aydin}.

Standard texts \cite{Morse,Bohren,Stratton} contain closed-form expressions of
$ \vec{M}_{smn}^{(j)}(k\#r)$ and $ \vec{N}_{smn}^{(j)}(k\#r)$   in spherical coordinates.
Whereas $ \vec{M}_{smn}^{(1)}(k\#r)$ and $ \vec{N}_{smn}^{(1)}(k\#r)$ are regular at the origin $r=0$, $ \vec{M}_{smn}^{(3)}(k\#r)$ and $ \vec{N}_{smn}^{(3)}(k\#r)$
are regular as $r\to\infty$.
 The  index $n$ denotes the order of the spherical Bessel function $j_n(kr)$ for $j=1$, and of the
 spherical Hankel functions $h_n^{(1)}(kr)$ for $j=3$, appearing in these wavefunctions. Furthermore, these wavefunctions involve the associated Legendre function $P_n^m(\cos \theta)$ of order $n$ and degree $m$,
and the index $s$ stands for either  even (e) or odd (o) parity. 

Closed-form expressions of the wavefunctions
$ \vec{M}_{smn}^{(j)}(k\=A^{-1}\.\#r)$ and $ \vec{N}_{smn}^{(j)}(k\=A^{-1}\.\#r)$ are also available
\cite{Lakhtakiaoriginal}. But those of $\vec{M}_{smn}^{(j)}(k \={A}^{-1} \.  \={S} \. \vec{r})$
and $\vec{N}_{smn}^{(j)}(k \={A}^{-1} \.  \={S} \. \vec{r})$ had to be derived for the work reported in this paper.

Let $\vec{r} \equiv \left( x, y , z \right) \equiv \left( r, \theta, \phi \right)$ and
$\vec{r}^\dprime \equiv \left( x^\dprime, y^\dprime , z^\dprime \right) \equiv \left( r^\dprime, \theta^\dprime, \phi^\dprime \right)$\, where
\begin{equation}
\vec{r}^\dprime= \={A}^{-1} \. \={S} \. \vec{r}\,.
\end{equation}
After expressing $\left( r^\dprime, \theta^\dprime, \phi^\dprime \right)$ in terms of $\left( r, \theta, \phi \right)$, as well as $\left( \ur^\dprime, \utheta^\dprime, \uphi^\dprime \right)$ in terms of $\left( \ur, \utheta, \uphi \right)$, one obtains
\begin{widetext}
\begin{equation} \label{m}
\begin{aligned}
\GerM^{(j)}_{smn} (\vec{r}) &=   \doubleunderline{R}_{cs}^T(\theta,\phi) \. \doubleunderline{S} \. \doubleunderline{A}^{-1} \. \doubleunderline{R}_{cs}(\theta,\phi) \. \bigg(
 \newJ_n^{(j)}(k\vec{r}) \big\{\ur \left[ t_{21}(\theta,\phi) \mathcal{Q}_{smn}(\theta,\phi) -t_{31}(\theta,\phi) \mathcal{R}_{smn}(\theta,\phi)\right] \\
& +\utheta \left[ t_{22}(\theta,\phi) \mathcal{Q}_{smn}(\theta,\phi)-t_{32}(\theta,\phi) \mathcal{R}_{smn}(\theta,\phi) \right]+ \uphi \left[ t_{23}(\theta,\phi) \mathcal{Q}_{smn}(\theta,\phi)-t_{33}(\theta,\phi) \mathcal{R}_{smn}(\theta,\phi)\right] \big\} \bigg ),
\end{aligned}
\end{equation}
and
\begin{equation} \label{n}
\begin{aligned}
\GerN^{(j)}_{smn} (\vec{r}) &=   \doubleunderline{R}_{cs}^T(\theta,\phi) \. \doubleunderline{S} \. \doubleunderline{A}^{-1} \.  \doubleunderline{R}_{cs}(\theta,\phi) \.\\
&\bigg( \ur \bigg\{t_{11}(\theta,\phi) \frac{\newJ_n^{(j)}(k\vec{r})}{k r g_1(\theta,\phi)} \mathcal{P}_{smn}(\theta,\phi)+ \newK_n^{(j)}(k\vec{r}) \left[ t_{21}(\theta,\phi) \mathcal{R}_{smn}(\theta,\phi)+t_{31}(\theta,\phi)\mathcal{Q}_{smn}(\theta,\phi) \right] \bigg\} \\
& + \utheta \bigg\{t_{12}(\theta,\phi) \frac{\newJ_n^{(j)}(k\vec{r})}{k r g_1(\theta,\phi)} \mathcal{P}_{smn}(\theta,\phi)+ \newK_n^{(j)}(k\vec{r}) \left[ t_{22}(\theta,\phi) \mathcal{R}_{smn}(\theta,\phi)+t_{32}(\theta,\phi)\mathcal{Q}_{smn}(\theta,\phi) \right] \bigg\}  \\
& +\uphi  \bigg\{t_{13}(\theta,\phi) \frac{\newJ_n^{(j)}(k\vec{r})}{k r g_1(\theta,\phi)} \mathcal{P}_{smn}(\theta,\phi)+ \newK_n^{(j)}(k\vec{r}) \left[ t_{23}(\theta,\phi) \mathcal{R}_{smn}(\theta,\phi)+t_{33}(\theta,\phi)\mathcal{Q}_{smn}(\theta,\phi) \right] \bigg\} \bigg),
\end{aligned}
\end{equation}
where the superscript $T$ denotes the transpose,
\begin{equation}
 \doubleunderline{R}_{cs}(\theta,\phi) = 
\begin{pmatrix}
\sin \theta \cos \phi &&  \cos \theta \cos \phi && -\sin \phi \\
\sin \theta \sin \phi&& \cos \theta \sin \phi && \cos \phi \\
\cos \theta && -\sin \theta && 0 
\end{pmatrix}\,,
\end{equation}
\begin{equation} \label{t11}
\begin{aligned}[b]
t_{11}(\theta,\phi) =& \frac{1}{g_1(\theta,\phi)}  \big[ g_2(\theta,\phi) (S_{32} \sin \theta \sin \phi+S_{31} \sin \theta \cos
   \phi+S_{33} \cos \theta)+ g_4(\theta,\phi)  (S_{12} \sin \theta
   \sin \phi+S_{11} \sin \theta \cos \phi  \\
   & + S_{13} \cos \theta
   )  + g_5(\theta,\phi)  (S_{22} \sin \theta \sin \phi+S_{21} \sin
   \theta \cos \phi+S_{23} \cos \theta) \big],
   \end{aligned}
\end{equation}
\begin{equation}
\begin{aligned}[b]
t_{12}(\theta,\phi) =&  \frac{1}{g_1(\theta,\phi)} \big[ g_2(\theta,\phi) (S_{31} \cos \theta \cos \phi+S_{32} \cos \theta \sin
   \phi-S_{33} \sin \theta)+ g_4(\theta,\phi)  (S_{11} \cos \theta
   \cos \phi + S_{12} \cos \theta \sin \phi \\
   & -S_{13} \sin\theta
   ) +g_5(\theta,\phi)  (S_{21} \cos \theta \cos \phi+S_{22} \cos
   \theta \sin \phi-S_{23} \sin \theta) \big],
   \end{aligned}
\end{equation}
\begin{equation}
\begin{aligned}[b]
t_{13}(\theta,\phi) =& \frac{1}{g_1(\theta,\phi)} \big[ -g_4(\theta,\phi)  (S_{11} \sin \phi-S_{12} \cos \phi)+g_5(\theta,\phi)
    (S_{22} \cos \phi-S_{21} \sin \phi)+g_2(\theta,\phi) (S_{32}
   \cos \phi-S_{31} \sin \phi) \big],
   \end{aligned}
\end{equation}
\begin{equation}
\begin{aligned}[b]
t_{21}(\theta,\phi) =& \frac{1}{{g_1(\theta,\phi) g_3(\theta,\phi)}} \big\{ g_2(\theta,\phi) \big[g_4(\theta,\phi)  (S_{12} \sin \theta \sin \phi+S_{11} \sin\theta \cos \phi+  S_{13} \cos \theta) +g_5(\theta,\phi)  (S_{22}
   \sin \theta \sin \phi \\
   &+ S_{21} \sin \theta \cos \phi +S_{23} \cos
   \theta)\big] -g_3^2(\theta,\phi) (S_{32} \sin \theta \sin \phi+S_{31}
   \sin \theta \cos \phi+S_{33} \cos \theta) \big\},
   \end{aligned}
\end{equation}
\begin{equation}
\begin{aligned}[b]
t_{22}(\theta,\phi) =& 
  \frac{1}{g_1(\theta,\phi) g_3(\theta,\phi)}\big\{ g_2(\theta,\phi) \big[ g_4(\theta,\phi)  (S_{11} \cos \theta \cos \phi+S_{12} \cos
   \theta \sin \phi-S_{13} \sin \theta) +g_5(\theta,\phi)  (S_{21}
   \cos \theta \cos \phi\\
   &+S_{22} \cos \theta \sin \phi -S_{23} \sin
   \theta)\big] -g_3^2(\theta,\phi) (S_{31} \cos \theta \cos \phi+S_{32}
   \cos \theta \sin \phi-S_{33} \sin \theta) \big\},
   \end{aligned}
\end{equation}
\begin{equation}
\begin{aligned}[b]
t_{23}(\theta,\phi) =& \frac{1}{g_1(\theta,\phi) g_3(\theta,\phi)} \big\{ g_2(\theta,\phi) \big[ g_5(\theta,\phi)  (S_{22} \cos \phi-S_{21} \sin \phi
   )-g_4(\theta,\phi)  (S_{11} \sin \phi-S_{12} \cos \phi
   )\big] \\
   &+g_3^2(\theta,\phi) (S_{31} \sin \phi-S_{32} \cos \phi ) \big\},
   \end{aligned}
\end{equation}
\begin{equation}
\begin{aligned}[b]
t_{31}(\theta,\phi) =& \frac{1}{g_3(\theta,\phi)} \big[ g_4(\theta,\phi)  (S_{22} \sin \theta \sin \phi+S_{21} \sin \theta
   \cos \phi+S_{23} \cos \theta) - g_5(\theta,\phi) (S_{12} \sin
   \theta \sin \phi+ S_{11} \sin \theta \cos \phi \\ 
   &+S_{13} \cos \theta
   ) \big],
   \end{aligned}
\end{equation}
\begin{equation}
\begin{aligned}[b]
t_{32}(\theta,\phi) =& \frac{1}{g_3(\theta,\phi)} \big[ g_4(\theta,\phi)  (S_{21} \cos \theta \cos \phi+S_{22} \cos \theta
   \sin \phi-S_{23} \sin \theta)-g_5(\theta,\phi)  (S_{11} \cos
   \theta \cos \phi+S_{12} \cos \theta  \sin \phi \\
   &-S_{13} \sin \theta
   ) \big],
   \end{aligned}
\end{equation}
and
\begin{equation} \label{t33}
t_{33}(\theta,\phi) = \frac{1}{g_3(\theta,\phi)} \big[ g_5(\theta,\phi)  (S_{11} \sin \phi-S_{12} \cos \phi)+g_4(\theta,\phi)
    (S_{22} \cos \phi-S_{21} \sin \phi) \big].
\end{equation}
In the foregoing expressions, $S_{\nu \nu^\prime}$, $\nu \in [1,3]$ and $\nu^\prime \in [1,3]$, are elements of the dyadic  $ \doubleunderline{S}$  written in matrix form as 
\begin{equation}
\begin{aligned}
 \doubleunderline{S}=
 \begin{pmatrix}
\cos \gamma \cos \beta \cos \alpha -\sin \gamma \sin \alpha && -\sin \gamma \cos \beta \cos \alpha-\cos \gamma \sin \alpha && \sin \beta \cos \alpha \\
 \cos \gamma \cos \beta \sin \alpha+ \sin \gamma \cos \alpha && -\sin \gamma \cos \beta \sin \alpha + \cos \gamma \cos \alpha && \sin \beta \sin \alpha \\
 -\cos \gamma \sin \beta  && \sin \gamma \sin \beta && \cos \beta
\end{pmatrix}.
\end{aligned}
\end{equation}
\end{widetext}
Furthermore, 
\begin{equation} \label{f1}
 \newJ_n^{(j)}(k\#r)= z_n[kr g_1(\theta,\phi)], 
 \end{equation}
 \begin{equation}\label{f2}
\newK_n(k\#r)= \frac{n+1}{krg_1(\theta,\phi)}\newJ_n^{(j)}(k\#r)-\newJ_{n+1}^{(j)}(k\#r),
 \end{equation}
  \begin{equation}\label{f3}
 \mathcal{Q}_{smn}(\theta,\phi)= \mathcal{Q}_{mn}(\theta,\phi) \mathcal{U}_{sm}(\theta,\phi), 
\end{equation}
  \begin{equation}\label{f4}
 \mathcal{R}_{smn}(\theta,\phi)= \mathcal{R}_{mn}(\theta,\phi) \mathcal{V}_{sm}(\theta,\phi),
 \end{equation}
  \begin{equation}\label{f5}
\mathcal{P}_{smn}(\theta,\phi)= \mathcal{P}_{mn}(\theta,\phi) \mathcal{V}_{sm}(\theta,\phi), 
\end{equation}
 \begin{equation}\label{f6}
 \mathcal{Q}_{mn}(\theta,\phi)= \frac{mg_1(\theta,\phi)}{g_3(\theta,
\phi)} P_n^m \left[ \frac{g_2(\theta,\phi)}{g_1(\theta,\phi)} \right],
\end{equation}
\begin{equation}\label{f7}
 \mathcal{P}_{mn}(\theta,\phi)= n(n+1)P_n^m \left[ \frac{g_2(\theta,\phi)}{g_1(\theta,\phi)} \right], 
\end{equation}
\begin{equation}\label{f8}
\begin{aligned}
& \mathcal{R}_{mn}(\theta,\phi)= \frac{g_1(\theta,\phi)}{g_3(\theta,\phi)} \bigg\{ (n-m+1) P_{n+1}^m \left[ \frac{g_2(\theta,\phi)}{g_1(\theta,\phi)} \right] \\
& -(n+1) \frac{g_2(\theta,\phi)}{g_1(\theta,\phi)} P_n^m \left[ \frac{g_2(\theta,\phi)}{g_1(\theta,\phi)} \right] \bigg\}, 
\end{aligned}
\end{equation}
\begin{equation}\label{f9}
 \calU_{sm}(\theta,\phi)=\lec\begin{array}{c}-\sin \left[mg_6(\theta,\phi)\right]\\\cos\left[mg_6(\theta,\phi)\right]\end{array}\ric
\,,
\quad s=\left\{\begin{array}{c}e\\{o}\end{array}\right.\,,
\end{equation}
and
\begin{equation}\label{f10}
 \calV_{sm}(\theta,\phi)=\lec\begin{array}{c}\cos \left[mg_6(\theta,\phi)\right]\\\sin\left[mg_6(\theta,\phi)\right]\end{array}\ric
\,,
\quad s=\left\{\begin{array}{c}e\\{o}\end{array}\right.\,,
\end{equation}
where $z_n(\.)$ is either $j_n(\.)$ when $j=1$, or $h_n^{(1)}(\.)$ when $j=3$. In Eqs. (\ref{t11})-(\ref{t33}) and (\ref{f1})-(\ref{f10}), the following angular functions have been used:
\begin{equation} \label{angular_function1}
g_1(\theta,\phi)= \left\{ \left[  g_2(\theta,\phi) \right]^2+ \left[  g_3(\theta,\phi)\right]^2 \right\}^{1/2}, 
\end{equation}
\begin{equation} \label{angular_function2}
   g_2(\theta,\phi)= \cos \beta  \cos \theta -\sin \beta  \sin \theta  \cos (\alpha +\phi ), 
   \end{equation}
   \begin{equation} \label{angular_function3}
   g_3(\theta,\phi)= \left\{ \left[  g_4(\theta,\phi) \right]^2+ \left[  g_5(\theta,\phi
   )\right]^2 \right\}^{1/2}, 
   \end{equation}
   \begin{equation} \label{angular_function4}
   \begin{aligned}
   g_4(\theta,\phi)=& \alpha_x \big\{ \cos \gamma  \big[ \cos \beta  \sin \theta  \cos (\alpha +\phi )+\sin \beta  \cos \theta \big] \\
   &-\sin \gamma\,  {\sin \theta}\, \sin (\alpha +\phi ) \big\}, 
   \end{aligned}
   \end{equation}
   \begin{equation} \label{angular_function5}
   \begin{aligned}
   g_5(\theta,\phi)=& \alpha_y \big\{ \sin \gamma  \big[ \cos \beta \sin \theta   \cos (\alpha +\phi ) +\sin \beta    \cos \theta \big] \\
   & + \cos \gamma\,  {\sin \theta}\, \sin (\alpha +\phi ) \big\}, 
   \end{aligned}
   \end{equation}
   and
   \begin{equation} \label{angular_function6}
   g_6(\theta,\phi)= \tan^{-1} \left[ \frac{g_5(\theta,\phi)}{g_4(\theta,\phi)} \right].
\end{equation}

\subsection{EBCM equations}
Consider an ellipsoid fully occupying the region $V$. Thus, $\#r(\theta,\phi)\in{V}\Rightarrow 
\vert\#r(\theta,\phi)\vert\le\vert\#r_S(\theta,\phi)\vert$. The region outside $V$ is vacuous.
With the stipulation that the source of the incident electromagnetic field lies outside
the sphere circumscribing $V$ \cite{Lakh-IskBook}, the incident electric field phasor may be
written as
\begin{equation}
 \label{1}
\begin{aligned}
\vec{E}_{\text{inc}}(\vec{r}) =&  \lim_{N\to\infty} \sum_{s \in \{e,o\}} \sum_{n=1}^N \sum_{m=0}^n  \big\{
D_{mn}\big[A_{smn}^{(1)} \vec{M}_{smn}^{(1)}(\ko \vec{r}) \\
& + B_{smn}^{(1)} \vec{N}_{smn}^{(1)}(\ko \vec{r}) \big]\big\}\,,
\end{aligned}
\end{equation}
where the normalization factor
\begin{equation}
D_{mn} = (2- \delta_{m0}) \frac{(2n+1)(n-m)!}{4n(n+1)(n+m)!}
\end{equation}
involves the Kronecker delta $\delta_{m m^\prime}$ and
 the expansion coefficients $A_{smn}^{(1)}$ and $B_{smn}^{(1)}$ are supposed to be  known
 $\,\forall\lec{s,m,n}\ric$.

Likewise, the scattered electric field phasor outside the sphere circumscribing $V$ can be expanded as
 \begin{equation}
 \label{2}
\begin{aligned}
\vec{E}_{\text{sca}}(\vec{r}) =&  \lim_{N\to\infty} \sum_{s \in \{e,o\}} \sum_{n=1}^N \sum_{m=0}^n
\big\{  D_{mn}\big[ A_{smn}^{(3)} \vec{M}_{smn}^{({3})}(\ko \vec{r}) \\
& + B_{smn}^{(3)} \vec{N}_{smn}^{({3})}(\ko \vec{r}) \big]\big\}\,,
\end{aligned}
\end{equation}
where the expansion coefficients $A_{smn}^{(3)}$ and $B_{smn}^{(3)}$  are not known.

Inside $V$, the electric and magnetic field phasors are represented by a superposition of the vector spherical wavefunctions derived in  Sec.~2.\ref{vector_wavefunctions} as
\begin{equation}
\vec{E}_{\text{int}}(\#r)=\lim_{N\to\infty}   \sum_{s\in\left\{e,o\right\}}\sum^{N}_{n=1}\sum^n_{m=0}
\left[
b_{smn}  \,\GerM^{(1)}_{smn}(\#r)+
c_{smn}  \,\GerN^{(1)}_{smn}(\#r)\right]\,
\label{Eint}
\end{equation}
and
\begin{eqnarray}
\nonumber
&&\vec{H}_{\text{int}}(\#r)=
-\frac{i}{\etao \eta_r}\lim_{N\to\infty}
  \sum_{s\in\left\{e,o\right\}}\sum^{N}_{n=1}\sum^n_{m=0}
\left[
b_{smn}  \,\GerN^{(1)}_{smn}(\#r)\right.
\\[5pt]
&&\qquad
+\left.
c_{smn}  \,\GerM^{(1)}_{smn}(\#r)\right]\,, 
\label{Hint}
\end{eqnarray}
where the expansion coefficients $b_{smn}$ and $c_{smn}$ are not known.

On (i) making use of the Ewald--Oseen extinction theorem and the
Huygens principle, (ii) exploiting the orthogonality properties of 
$\vec{M}_{smn}^{({j})}(\ko \vec{r})$ and
$\vec{N}_{smn}^{({j})}(\ko \vec{r})$ on a unit sphere \cite{Lakhtakiaoriginal}, and (iii) application of the  continuity of tangential electric and magnetic fields across $S$, a matrix relation emerges between the incident field and scattered field coefficients \cite{Waterman,Lakh-IskBook}. This relation is written compactly as
\begin{equation}
\begin{bmatrix}
\les{A^{(3)}}\ris \\ --- \\ \les{ B^{(3)}}\ris
\end{bmatrix}
=[T]
\begin{bmatrix}
\les{A^{(1)}}\ris \\ --- \\ \les{ B^{(1)}}\ris
\end{bmatrix}\,,
\label{A3B3A1B1}
\end{equation}
where the column vectors $\les{A^{(j)}}\ris$ 
and $\les{B^{(j)}}\ris$ contain the coefficients $A^{(j)}_{smn}$
and $B^{(j)}_{smn}$, respectively.

The matrix
\begin{equation} 
\label{4a}
[T]= -[Y^{(3)}] [Y^{(1)}]^{-1}
\end{equation}
is called the transition matrix or the T matrix.
The matrix $[Y^{(j)}]$, $j\in[1,3]$, is symbolically written as
\begin{equation} 
\label{5}
[Y^{(j)}]= 
\begin{pmatrix}
I_{smn,s^\prime m^\prime n^\prime}^{(j)} && \big| && J_{smn,s^\prime m^\prime n^\prime}^{(j)} \\
---- && \big| && ---- \\
K_{smn,s^\prime m^\prime n^\prime}^{(j)}&& \big| && L_{smn,s^\prime m^\prime n^\prime}^{(j)}
\end{pmatrix}.
\end{equation}

The matrix elements in Eq. (\ref{5}) are double integrals given by 
\begin{equation} 
\label{6-I}
\begin{aligned}[b]
I_{smn,s^\prime m^\prime n^\prime}^{(j)} =& -\frac{i \ko^2}{\pi} 
 \iint \limits_{\calS} \,d^2r_s \\
& \Bigg[
\lec\vec{N}_{smn}^{(\ell)}(\ko \vec{r}_s) \. 
\les \nhat(\vec{r}_s)  \times  \GerM^{(1)}_{s^\prime m^\prime n^\prime}(\vec{r}_s)  \ris\ric
\\
+ &\eta_r^{-1}
\lec\vec{M}_{smn}^{(\ell)}(\ko \vec{r}_s) \. 
\les \nhat(\vec{r}_s)  \times  \GerN^{(1)}_{s^\prime m^\prime n^\prime}(\vec{r}_s) \ris\ric 
\Bigg]
\end{aligned}
\end{equation}
\begin{equation} 
\label{6-J}
\begin{aligned}[b]
J_{smn,s^\prime m^\prime n^\prime}^{(j)} =& -\frac{i \ko^2}{\pi} 
 \iint \limits_{\calS} \,d^2r_s \\
& \Bigg[
\lec\vec{N}_{smn}^{(\ell)}(\ko \vec{r}_s) \. 
\les \nhat(\vec{r}_s)  \times  \GerN^{(1)}_{s^\prime m^\prime n^\prime}(\vec{r}_s)  \ris\ric
\\
+ &\eta_r^{-1}
\lec\vec{M}_{smn}^{(\ell)}(\ko \vec{r}_s) \. 
\les \nhat(\vec{r}_s)  \times  \GerM^{(1)}_{s^\prime m^\prime n^\prime}(\vec{r}_s) \ris\ric 
\Bigg]
\end{aligned}
\end{equation}
\begin{equation} 
\label{6-K}
\begin{aligned}[b]
K_{smn,s^\prime m^\prime n^\prime}^{(j)} =& -\frac{i \ko^2}{\pi} 
 \iint \limits_{\calS} \,d^2r_s \\
& \Bigg[
\lec\vec{M}_{smn}^{(\ell)}(\ko \vec{r}_s) \. 
\les \nhat(\vec{r}_s)  \times  \GerM^{(1)}_{s^\prime m^\prime n^\prime}(\vec{r}_s)  \ris\ric
\\
+ &\eta_r^{-1}
\lec\vec{N}_{smn}^{(\ell)}(\ko \vec{r}_s) \. 
\les \nhat(\vec{r}_s)  \times  \GerN^{(1)}_{s^\prime m^\prime n^\prime}(\vec{r}_s) \ris\ric 
\Bigg]
\end{aligned}
\end{equation}
and
\begin{equation} 
\label{6-L}
\begin{aligned}[b]
L_{smn,s^\prime m^\prime n^\prime}^{(j)} =& -\frac{i \ko^2}{\pi} 
 \iint \limits_{\calS} \,d^2r_s \\
& \Bigg[
\lec\vec{M}_{smn}^{(\ell)}(\ko \vec{r}_s) \. 
\les \nhat(\vec{r}_s)  \times  \GerN^{(1)}_{s^\prime m^\prime n^\prime}(\vec{r}_s)  \ris\ric
\\
+ &\eta_r^{-1}
\lec\vec{N}_{smn}^{(\ell)}(\ko \vec{r}_s) \. 
\les \nhat(\vec{r}_s)  \times  \GerM^{(1)}_{s^\prime m^\prime n^\prime}(\vec{r}_s) \ris\ric 
\Bigg]
\end{aligned}
\end{equation}
In the foregoing double integrals, $\nhat(\vec{r}_s)$
is the unit outward normal to $\calS$ at $\vec{r}_s \in \calS$ and $\ell=j+2( \text{mod} 4) \in [3,1]$.

\subsection{Plane-wave incidence}
Consider a plane wave   incident on the ellipsoid such  that
\begin{equation}
\label{Einc}
\vec{E}_{\text{inc}}(\vec{r})= \e_{\text{inc}} \exp(i\ko\k_{\text{inc}} \. \vec{r}),
\end{equation}
where the unit vector
\begin{equation}
\k_{\text{inc}}=  (\ux \sin \theta_{\text{inc}} \cos \phi_{\text{inc}} + \uy \sin \theta_{\text{inc}} \sin \phi_{\text{inc}}+\uz \cos \theta_{\text{inc}})
\end{equation}
contains the angles  $\theta_{\text{inc}}\in[0,\pi]$ and $\phi_{\text{inc}}\in[0,2\pi)$ that together define the incidence direction.  

The unit vector $\e_{\text{inc}}$ defines the polarization state, 
with the stipulation that $\k_{\text{inc}}\.\e_{\text{inc}}=0$.
For any specific choice of  $\k_{\text{inc}}$, a choice has to be made for $\e_{\text{inc}}$.
If this choice is denoted by $\e\one_{\text{inc}}$, the second choice becomes automatic:
$\e\two_{\text{inc}}= \k_{\text{inc}}\times\e\one_{\text{inc}}$. 
Thus,
\begin{equation}
\label{def-ep}
\ep_{\text{inc}}=  \ux \cos \theta_{\text{inc}} \cos \phi_{\text{inc}}  + \uy \cos \theta_{\text{inc}} \sin \phi_{\text{inc}} -\uz \sin \theta_{\text{inc}}
\end{equation}
and
\begin{equation}
\label{def-es}
\es_{\text{inc}}=  -\ux \sin \phi_{\text{inc}}  + \uy \cos \phi_{\text{inc}}\,
\end{equation}
represent the parallel and perpendicular polarization states, respectively; likewise,
\begin{equation}
\label{def-eincLCP}
\eLCP_{\text{inc}}=  \frac{1}{\sqrt{2}}\left(\ep_{\text{inc}} +i\es_{\text{inc}}\right)
\end{equation}
and
\begin{equation}
\label{def-eincRCP}
\eRCP_{\text{inc}}=  \frac{1}{\sqrt{2}}\left(\ep_{\text{inc}} -i\es_{\text{inc}}\right)\,,
\end{equation}
respectively,
represent the left-circular and right-circular polarization (LCP and RCP) states.

The expansion coefficients on the right side of Eq.~(\ref{1}) are given by \cite{Morse,Alkhoori1}
\begin{equation} 
\label{incident_expansion}
\begin{rcases}
\begin{aligned}[b]
A_{smn}^{(1)} = 4 i^n \sqrt{n(n+1)}\, \e_{\text{inc}} \. \vec{C}_{smn}(\theta_{\text{inc}},\phi_{\text{inc}})  \\
B_{smn}^{(1)} = 4 i^{n -1} \sqrt{n(n+1)}\,  \e_{\text{inc}}\.\left[\k_{\text{inc}} \times \vec{C}_{smn}(\theta_{\text{inc}},\phi_{\text{inc}})\right] 
\end{aligned}
\end{rcases}\,,
\end{equation}
where the vector spherical harmonic 
\begin{eqnarray}
\nonumber
&&
 \vec{C}_{smn}(\theta,\phi)=
  \frac{1}{\sqrt{n(n+1)}} \left[ \mp m \frac{P_n^m(\cos \theta)}{\sin \theta}  \Bigg\{ \begin{matrix}
\sin (m \phi)\\
\cos (m \phi)
\end{matrix} \Bigg\} \utheta \right.
\\[8pt]
&& \quad\left.-\frac{d P_n^m(\cos \theta)}{d \theta} \Bigg\{ \begin{matrix}
\cos (m\phi)\\
\sin (m\phi)
\end{matrix} \Bigg\}  \uphi  \right]  \,,\quad
s=\left\{\begin{array}{l} e \\ o\end{array}\right.\,.
\end{eqnarray}

Let the expansion coefficients $A\smn\oneparl$ and $B\smn\oneparl$ represent the
incident parallel-polarized plane wave, 
whereas the expansion coefficients $A\smn\oneperp$ and $B\smn\oneperp$ the
incident perpendicularly polarized plane wave. Then, the  twin 
relations
\begin{equation}
\left.\begin{array}{l}
A\smn\oneparl = i B\smn\oneperp
\,,\qquad
B\smn\oneparl = i A\smn\oneperp
\end{array}\right\}
\label{rel-parl-perp}
\end{equation}
emerge from Eqs.~(\ref{def-ep}), (\ref{def-es}), and (\ref{incident_expansion}).

Let the expansion coefficients $A\smn\oneLCP$ and $B\smn\oneLCP$ represent the
incident LCP plane wave, whereas $A\smn\oneRCP$ and $B\smn\oneRCP$ repesent the
incident RCP plane wave. Then, 
\begin{equation}
\left.\begin{array}{l}
A\smn\oneLCP=  \frac{1}{\sqrt{2}}\left( A\smn\oneparl +i A\smn\oneperp\right)
\\[5pt]
B\smn\oneLCP=  \frac{1}{\sqrt{2}}\left( B\smn\oneparl +i B\smn\oneperp\right)
\end{array}\right\}
\end{equation}
and
\begin{equation}
\left.\begin{array}{l}
A\smn\oneRCP=  \frac{1}{\sqrt{2}}\left( A\smn\oneparl -i A\smn\oneperp\right)
\\[5pt]
B\smn\oneRCP=  \frac{1}{\sqrt{2}}\left( B\smn\oneparl -i B\smn\oneperp\right)
\end{array}\right\}\,,
\end{equation}
follow from Eqs.~(\ref{def-eincLCP}) and (\ref{def-eincRCP}). Using Eqs.~(\ref{rel-parl-perp}) 
in the foregoing relations finally yields the twin relations 
\begin{equation}
\left.\begin{array}{l}
A\smn\oneLCP=   B\smn\oneLCP
\,,\qquad
A\smn\oneRCP= -B\smn\oneRCP
\end{array}\right\}\,.
\label{rel-LCP-RCP}
\end{equation}

\subsection{Total scattering, extinction, and absorption efficiencies}\label{sae}
In the far zone, the scattered electric field phasor 
can be approximated as \cite{Bowman,Bohren}
\begin{equation}
\vec{E}_{\text{sca}}(r,\theta,\phi) \approx \vec{F}_{\text{sca}}( \theta,\phi) \,{\exp(i\ko r)}/{r}
\end{equation}
where \cite{Alkhoori1}
\begin{equation}
\begin{aligned}[b]
\vec{F}_{\text{sca}}(\theta,\phi)=& 
\left({1}/{\ko}\right)
\lim_{N\to\infty}
 \sum_{s \in \{e,o\}} \sum_{n=1}^{N} \sum_{m=0}^n \bigg\{(-i)^n D_{mn} \sqrt{n(n+1)} \\
& \big[ -i A_{smn}^{(3)} \={I} + B_{smn}^{(3)}\ur \times \={I}   \big]\. \vec{C}_{smn}(\theta,\phi) \bigg\}.
\end{aligned}
\label{def-Fsca}
\end{equation}
The differential scattering efficiency is defined as
 \begin{equation}
\label{sigmaD-def}
\QD(\theta,\phi)= \left({4}/{c^2}\right)  { \vert\vec{F}_{\text{sca}}(\theta,\phi)\vert^2}
 \,.
\end{equation}
Upon integrating $\QD(\theta,\phi)$ over the entire solid angle, the total scattering efficiency is obtained as 
\begin{eqnarray}
\nonumber
&&
\Qsca   =  {\left(\ko c\right)^{-2}}  
\\
&&\quad\times \lim_{N\to\infty}
\sum_{s \in \{e,o\}} \sum_{n=1}^{N} \sum_{m=0}^n  D_{mn} \left[ \vert{A_{smn}^{(3)}}\vert^2+ \vert{B_{smn}^{(3)}}\vert^2 \right]\,.
\label{def-Qsca}
\end{eqnarray}

By virtue of the forward-scattering theorem \cite{Karam1982,Baum2007},
the extinction efficiency can be calculated as
\begin{equation}
\label{def-Qext}
\Qext=
\left({4}/{\ko c^2}\right) 
{\rm Im}\les
\vec{F}_{\text{sca}}(\theta\inc,\phi\inc)\. \e_{\text{inc}}^\ast
\ris \,,
\end{equation}
where the asterisk denotes the complex conjugate.
Substitution of
Eqs.~(\ref{incident_expansion}) and (\ref{def-Fsca}) on the
right side of Eq.~(\ref{def-Qext})
yields
\begin{eqnarray}
\nonumber
&&
\Qext=    {\left(\ko c\right)^{-2}}  
\\
&&
\nonumber
\times
\lim_{N\to\infty}
 \sum_{s \in \{e,o\}} \sum_{n=1}^{N} \sum_{m=0}^n
 D_{mn} {\rm Re} \left[
 -A_{smn}^{(3)}A_{smn}^{(1)\ast}+  B_{smn}^{(3)}B_{smn}^{(1)\ast}
 \right]\,.
 \\
 &&
\end{eqnarray}
The absorption efficiency can then be calculated as
\begin{equation}
\label{def-Qabs}
\Qabs=\Qext-\Qsca\,.
\end{equation}

The scattered field coefficients $A\smn\threeparl$ and $B\smn\threeparl$
are related to the incident field coefficients $A\smn\oneparl$ and $B\smn\oneparl$,
and so on, through Eq.~(\ref{A3B3A1B1}). Accordingly,
\begin{equation}
\left.\begin{array}{l}
A\smn\threeLCP=  \frac{1}{\sqrt{2}}\left( A\smn\threeparl +i A\smn\threeperp\right)
\\[5pt]
B\smn\threeLCP=  \frac{1}{\sqrt{2}}\left( B\smn\threeparl +i B\smn\threeperp\right)
\end{array}\right\}
\end{equation}
and
\begin{equation}
\left.\begin{array}{l}
A\smn\threeRCP=  \frac{1}{\sqrt{2}}\left( A\smn\threeparl -i A\smn\threeperp\right)
\\[5pt]
B\smn\threeRCP=  \frac{1}{\sqrt{2}}\left( B\smn\threeparl -i B\smn\threeperp\right)
\end{array}\right\}\,.
\end{equation}
Equations~(\ref{def-Qsca}), (\ref{def-Qext}), and (\ref{def-Qabs}) 
then deliver the identities
\begin{equation}
\left.\begin{array}{l}
\QscaLCP+\QscaRCP= \Qscaparl +\Qscaperp
\\[5pt]
\QextLCP+\QextRCP= \Qextparl +\Qextperp
\\[5pt]
\QabsLCP+\QabsRCP= \Qabsparl +\Qabsperp
\end{array}\right\}\,.
\label{rel-CP-LP}
\end{equation}
These identities hold regardless of the composition of the scattering material, so long as it is linear.

\subsection{ Impedance-matched scattering material }  \label{impmatch}
When $\etar=1$, i.e., $\epsr=\mur$, the   material chosen for the scatterer can be said to be impedance matched
to free space. The equalities
 $I_{smn,s^\prime m^\prime n^\prime}^{(j)}=L_{smn,s^\prime m^\prime n^\prime}^{(j)}$ and $J_{smn,s^\prime m^\prime n^\prime}^{(j)}=K_{smn,s^\prime m^\prime n^\prime}^{(j)}$ then follow
 from Eqs. (\ref{6-I})-(\ref{6-L}). In consequence, if the
 the  column vectors $[A^{(1)}]$ and $[B^{(1)}]$ on the right side of Eq.~(\ref{A3B3A1B1})
 were to be interchanged,
 then  the  column vectors $[A^{(3)}]$ and $[B^{(3)}]$ on the right side of Eq.~(\ref{A3B3A1B1})
 shall have to be interchanged as well;  both $\QD(\theta,\phi)$ and $\Qsca$ would remain
 unaffected because $\ur \. \vec{C}_{smn}(\theta,\phi)\equiv 0$ \cite{Alkhoori2}.
Consequently,   both $\QD (\theta, \phi)$ and $\Qsca$ do not depend on the polarization
 state of the incident plane wave. Additional analysis \cite{Alkhoori2} shows that
 $\Qabs$ and $\Qext$ also do not depend on the polarization
 state of the incident plane wave.

\section{Numerical Results and Discussion} \label{s2}
The T matrix was computed on the 
Mathematica\texttrademark~version 11.3 platform implemented
on a Dell Alienware 18 laptop computer with 16~GB memory.
We used the Gauss--Legendre quadrature scheme
\cite{Jaluria} to evaluate the integrals in Eqs. (\ref{6-I})-(\ref{6-L}). 
By testing against known integrals \cite{GS} so as to compute  integrals correct to $\pm0.1\%$ relative error,  the number of nodes was chosen to be $24$ for integration over $\theta$ as well as for integration over $\phi$.
 The computation of the inverse of $[Y^{(1)}]$ was carried out using the lower-upper decomposition  method \cite{math}. The value of $N$
was incremented by unity until the backscattering efficiency
$\Qb=\QD(-\k_{\text{inc}})$ converged within a tolerance of $\pm0.1\%$.

Validation of the program was accomplished by comparing with results available for simpler problems, such as Lorenz--Mie theory for isotropic dielectric/magnetic/dielectric-magnetic spheres \cite{Stratton, Bohren}. For a sphere as well as for an ellipsoid composed of a material described by Eqs.~(\ref{D})--(\ref{mata}) with $\={S}=\={I}$, our program was completely in accord with published data \cite{Jafri,Jafri-err, Alkhoori1}.  

\begin{figure}[htb]
 \centering 
     \begin{subfigure}[h]{0.8\textwidth}
\includegraphics[width=0.8\linewidth]{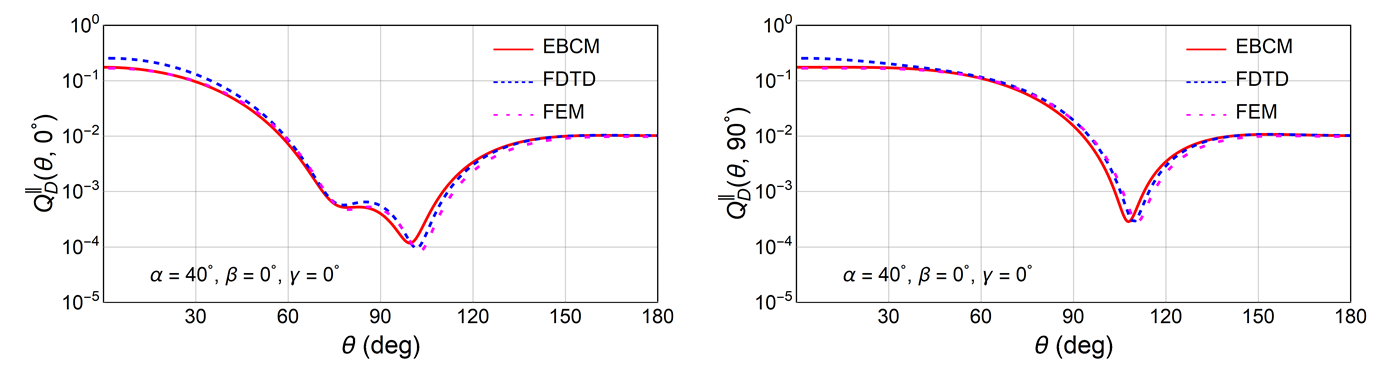}
 \end{subfigure}
\caption{{(Left) $\QD^\parallel(\theta, 0^\circ)$ and (right)  $\QD^\parallel(\theta, 90^\circ)$ vs. $\theta \in [0^\circ, 180^\circ]$ for a biaxially dielectric-magnetic ellipsoid described by $\epsr=2$, $\mur=1.05$, $\alpha_x=1.2$, $\alpha_y=1.1$, $\alpha=40^\circ$, $\beta=\gamma=0^\circ$, $a/c=1/2$, $b/c=2/3$, and $\koc =3$, when   $\theta_{\text{inc}}=0^\circ$ and $\phi_{\text{inc}}=0^\circ$.  }}
\label{rot_z}
\end{figure}

\begin{figure}[htb]
 \centering 
     \begin{subfigure}[h]{0.8\textwidth}
\includegraphics[width=0.8\linewidth]{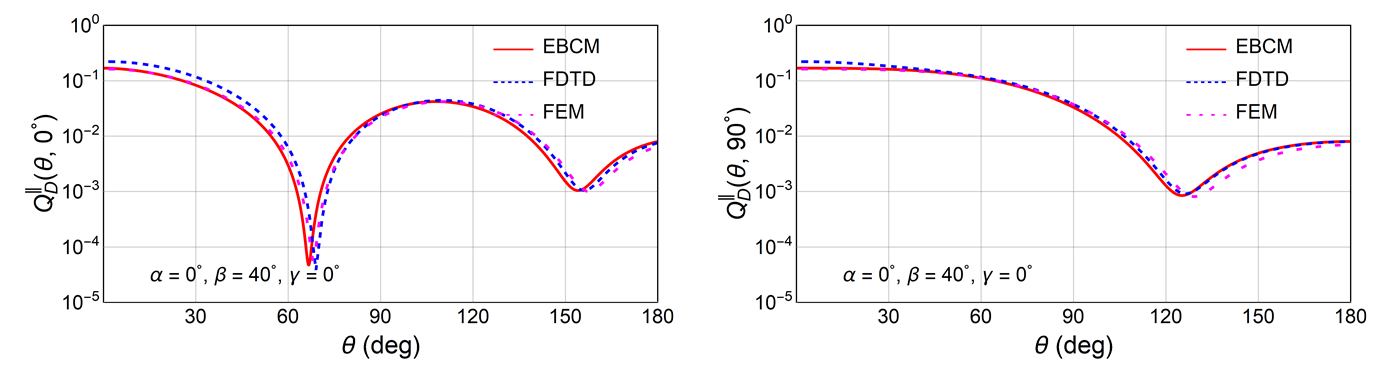}
 \end{subfigure}
\caption{{Same as Fig. \ref{rot_z}, except that $\beta = 40^\circ$ and $\alpha = \gamma =0^\circ$.}}
\label{rot_y}
\end{figure}

{When $\={S} \neq \={I}$, our program was validated by comparing its results against results obtained by FEKO\texttrademark~software \cite{FEKO}. The FEKO results were based on the 
finite-difference time-domain (FDTD) method \cite{FDTD_book} and the finite-element method (FEM) \cite{Monk_book}. Figure \ref{rot_z} shows plots of $\QD^\parallel(\theta, 0^\circ)$ and $\QD^\parallel(\theta, 90^\circ)$ vs. $\theta \in [0^\circ, 180^\circ]$ for a biaxially dielectric-magnetic ellipsoid described by $\epsr=2$, $\mur=1.05$, $\alpha_x=1.2$, $\alpha_y=1.1$, $\alpha=40^\circ$, $\beta=\gamma=0^\circ$, $a/c=1/2$, $b/c=2/3$, and $\koc =3$. The direction of incidence is specified by $\theta_{\text{inc}}=0^\circ$ and $\phi_{\text{inc}}=0^\circ$. We see a good agreement among the EBCM, FDTD and FEM results. We repeated these calculations for Fig. \ref{rot_y}, but with $\beta = 40^\circ$ and $\alpha = \gamma =0^\circ$. Again, good agreement among the EBCM, FDTD, and FEM data is evident.}

 \begin{table} 
\begin{center}
\caption{\label{Table1}
			{\bf  Adequate value of $N$ and computation time on a Dell Alienware 18 laptop
			computer to determine $\Qb$ and $\Qsca$ for a sphere or an ellipsoid ($a/c=1/2$
and $b/c=2/3$),  when $\epsr=2$, $\mur=1.05$,   $\koc=3$, $\theta_{\text{inc}}=45^\circ$,
 $\phi_{\text{inc}}=30^\circ$, and  $\e_{\text{inc}}=\ep_{\text{inc}}$.
 }}
 \begin{tabular}{|c| c |c |c| c|}
 \hline
 Object &  Anisotropy &  Orientation & $N$ & Computation  \\ [0.5ex] 
   &  parameters &  angles & & time (s)  \\
 \hline\hline
Ellipsoid  
&$\alpha_x= 1.2$, $\alpha_y =1.1$
& $\alpha= 20^\circ$, $\beta = 40^\circ$, $\gamma =30^\circ$
& $8$ 
& $8940.44$  
\\
\hline
  Ellipsoid  
  &$\alpha_x= 1.2$, $\alpha_y =1.1$ 
  & $\alpha=\beta = \gamma =0^\circ$ 
  & $7$ 
  & $4991.35$  
  \\ 
  \hline
 Ellipsoid  
 &$\alpha_x= \alpha_y =1$ 
 &$\alpha=\beta = \gamma =0^\circ$ 
 & $6$ 
 & $2617.74$  
 \\ 
\hline

Sphere 
& $\alpha_x= 1.2$, $\alpha_y =1.1$
& $\alpha= 20^\circ$, $\beta = 40^\circ$, $\gamma =30^\circ$ 
& $7$ 
& $6408.86$ 
\\ 
\hline 
Sphere 
& $\alpha_x= 1.2$, $\alpha_y =1.1$
& $\alpha=\beta = \gamma =0^\circ$
& $6$ 
& $3034.59$  
\\ 
\hline

Sphere 
& $\alpha_x= \alpha_y =1$
& $\alpha=\beta = \gamma =0^\circ$
& $5$ 
& $1593.66$  

\\ 
\hline
Sphere (using symmetries 
& $\alpha_x= \alpha_y =1$
& $\alpha=\beta = \gamma =0^\circ$
& $5$ 
& $56.94$  
\\ 
of double integrals) 
& 
& 
& 
&   
\\
 \hline
\end{tabular}
\end{center} 
\end{table} 

Anisotropy considerably enhances computational requirements, as summarized in Table~\ref{Table1}. For example,
$N=7$ is needed for a sphere ($a/c=b/c=1$) and $N = 8$ for an ellipsoid ($a/c=1/2$
and $b/c=2/3$), when we choose $\epsr=2$, $\mur=1.05$, $\alpha_x=1.2$, $\alpha_y=1.1$, $\alpha=20^\circ$, $\beta=40^\circ$, $\gamma=30^\circ$,  $\koc=3$, $\theta_{\text{inc}}=45^\circ$,
and $\phi_{\text{inc}}=30^\circ$. Because $\epsr$ and $\mur$ are both real, we
exploited the relation $[Y^{(3)}] = {\rm Re}[Y^{(1)}]$ in order to reduce computation time. 
When
we set $\alpha=\beta=\gamma=0^\circ$ so that the constitutive
principal axes coincide with the shape principal axes,  
$N=6$ is needed for the  sphere and $N=7$ for the  ellipsoid. However, we need $N=5$
and $N=6$ for the sphere
and the ellipsoid, respectively,
when we eliminate anisotropy by setting $\alpha_x=\alpha_y=1$.
Another way to appreciate the enhancement of computational requirements by anisotropy is
to examine the computation time. According to Table~\ref{Table1}, anisotropy  doubles
the computation time even when the constitutive
principal axes coincide with the shape principal axes; the computation time increases
significantly when the two sets of principal axes do not coincide.  Incidentally,
the double integrals defined in Eqs.~(\ref{6-I})--(\ref{6-L})
simply considerably for an isotropic 
dielectric-magnetic spheroid  \cite{Barberbook}
due to   rotational symmetry about the $z$ axis (i.e.,
$r_s$ is independent of $\phi$),
and even more simplification occurs for an isotropic 
dielectric-magnetic sphere because  $r_s =c$ then for all values of $\theta$ and $\phi$
in Eq.~(\ref{rs-def}).  When incorporated in a computer program, such simplifications 
greatly reduce the computation time, as can be deduced by comparing the last two rows
of Table~\ref{Table1}. Also, we note that concurrent computation of a substantial number
of double integrals using multiple  processors in parallel will reduce the computation time.

Now, we present numerical results on the total scattering efficiency of a biaxial dielectric-magnetic ellipsoid  in relation to 
\begin{itemize}
\item  the  propagation direction $\k_{\text{inc}}$  of the incident plane wave;
\item the polarization state of the incident plane wave ($\e_{\text{inc}}$);
\item  the constitutive-anisotropy parameters  $\alpha_x$ and $\alpha_y$; 
\item  the nonsphericity parameters $a/c$ and $b/c$;
\item the orientation angles $\alpha$, $\beta$, and $\gamma$; and
\item the electrical length $\koc$.
\end{itemize} 
Comparison is made to analogous numerical results for 
a biaxial dielectric-magnetic spheroid and a biaxial dielectric-magnetic sphere.

\subsection{Orientational mismatch of $\=C$ and $\=U$ \label{sec3A}}

\begin{figure}[htb]
 \centering 
     \begin{subfigure}[h]{0.4\textwidth}
\includegraphics[width=0.8\linewidth]{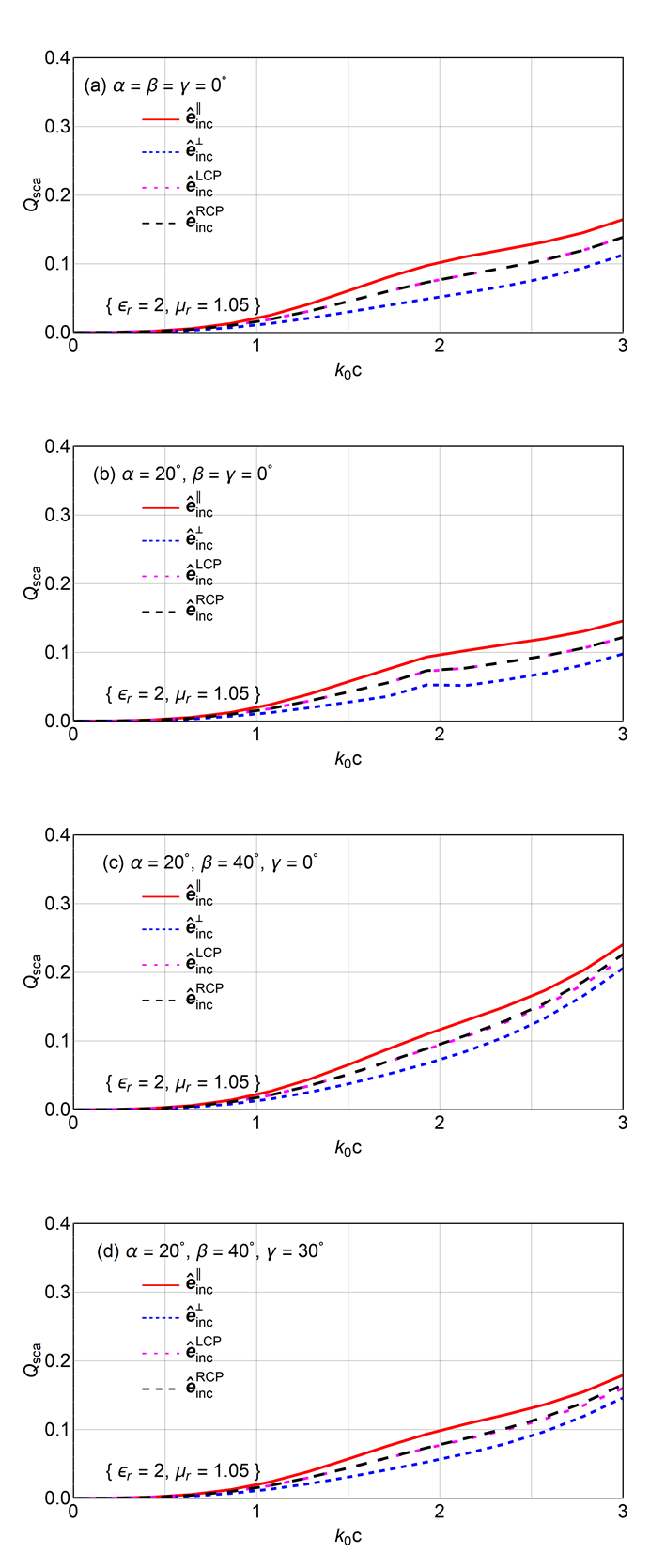}
 \end{subfigure}
\caption{ $\Qscaparl$, $\Qscaperp$, $\QscaLCP$,
and $\QscaRCP$ vs. $\koc\in[0,3]$ for  a biaxially dielectric-magnetic ellipsoid described by $\epsr=2$, $\mur=1.05$, $\alpha_x=1.2$, $\alpha_y=1.1$, $a/c=1/2$, and $b/c=2/3$. The angles of incidence are $\theta_{\text{inc}}=45^\circ$ and $\phi_{\text{inc}}=30^\circ$.  
(a) $\alpha=\beta=\gamma=0^\circ$; (b) $\alpha=20^\circ$ and $\beta=\gamma=0^\circ$; (c) $\alpha=20^\circ$, $\beta=40^\circ$, and $\gamma=0^\circ$; and (d) $\alpha=20^\circ$, $\beta=40^\circ$, and $\gamma=30^\circ$.} 
\label{Qsca_general}
\end{figure}

In order to illustrate the effects of the lack of coincidence of the principal axes of $\=C$
and $\=U$, Fig.~\ref{Qsca_general} shows plots of 
$\Qscaparl$, $\Qscaperp$, $\QscaLCP$,
and $\QscaRCP$   vs. $\koc \in [0,3]$ for a biaxially dielectric-magnetic ellipsoid, described by
$\epsr=2$, $\mur=1.05$, $\alpha_x=1.2$, $\alpha_y=1.1$, $a/c=1/2$, and $b/c=2/3$.
The incidence direction is specified by $\theta_{\text{inc}}=45^\circ$ and $\phi_{\text{inc}}=30^\circ$.
Since   the scattering material is non-dissipative,
$\Qabs\equiv0$ for all polarization states of the incident plane wave.

Figure~\ref{Qsca_general}(a) depicts plots of 
$\Qscaparl$, $\Qscaperp$, $\QscaLCP$,
and $\QscaRCP$ vs. $\koc$ when the principal axes of  $\=C$ and $\=U$ coincide, i.e.,
$\alpha=\beta=\gamma=0^\circ$. We see that $\Qscaparl>\Qscaperp$   regardless of $\koc \in [0,3]$. Furthermore, $\QscaLCP\simeq\QscaRCP$, so that
$\Qsca$ for either circularly polarized plane wave is almost the arithmetic mean of  $\Qscaparl$ and $\Qscaperp$ by virtue of Eq.~(\ref{rel-CP-LP})$_1$.

Rotation of $\=C$ about the $z$ axis of the laboratory coordinate system has a small but noticeable effect on $\Qsca$,
as becomes evident from comparing Fig.~\ref{Qsca_general}(a) with Fig.~\ref{Qsca_general}(b),
the latter being drawn for $\alpha=20^\circ$ and $\beta=\gamma=0^\circ$. A subsequent rotation by 
$\beta=40^\circ$ about the new $y$ axis also affects $\Qsca$, regardless of the polarization state
of the incident plane wave;
see  Fig.~\ref{Qsca_general}(c) drawn for $\alpha=20^\circ$,  $\beta=30^\circ$, and $\gamma=0^\circ$. 
The final rotation by $\gamma=30^\circ$ about the latest $z$ axis also has an effect on $\Qsca$ in
Fig.~\ref{Qsca_general}(d), regardless of the polarization state of the incident plane wave and the electrical size of the ellipsoid. In all panels of Fig.~\ref{Qsca_general}, for the most part $\Qsca$  increases
as $\koc$ does.  The excess of $\QscaRCP$ over $\QscaLCP$ is  clearer in Figs. \ref{Qsca_general}(c,d) than in  Figs.~\ref{Qsca_general}(a,b).

\begin{figure}[h]
 \centering 
     \begin{subfigure}[h]{0.4\textwidth}
\includegraphics[width=0.8\linewidth]{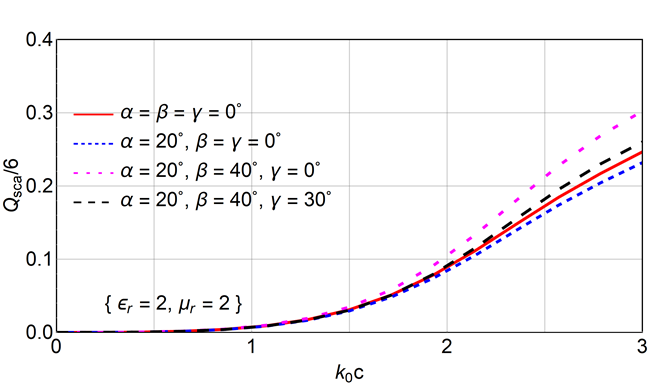}
 \end{subfigure}
\caption{$\Qsca$ vs. $\koc\in[0,3]$ for  a biaxially dielectric-magnetic ellipsoid described by $\epsr=\mur=2$, $\alpha_x=1.2$, $\alpha_y=1.1$, $a/c=1/2$, and $b/c=2/3$. The angles of incidence are $\theta_{\text{inc}}=45^\circ$ and $\phi_{\text{inc}}=30^\circ$.}
\label{Qsca_general_matched}
\end{figure}

 Figure~\ref{Qsca_general_matched} depicts plots of  $\Qsca$ vs. $\koc$ for the same biaxially dielectric-magnetic ellipsoid of Fig.~\ref{Qsca_general}, except that $\epsr=\mur=2$. Since the scattering material is impedance matched to the surrounding free space,   $\Qsca$ is independent of $\e_{\text{inc}}$ as   discussed in Sec.~\ref{s1}.\ref{impmatch}. The effect of  orientational mismatch of $\=C$ and $\=U$  as $\koc$ increases is readily apparent  in this figure, and the effects of  various rotations  are exactly the same as those reported for Fig.~\ref{Qsca_general}.
 
 Next, we consider scattering materials with: 
\begin{itemize}
\item[(a)] $\epsr\mur<0 \Rightarrow\lec\epsr\gtrless0,\mur\lessgtr0\ric$ 
and
\item [(b)]    $\epsr\mur>0 \Rightarrow\lec\epsr\gtrless0,\mur\gtrless0\ric$,
\end{itemize}
both $\epsr$ and $\mur$ being purely real.
Simultaneous reversal of the signs of $\epsr$ and $\mur$ in case (a)  changes neither
the relative wavenumber $k/\ko$ nor the relative impedance $\etar$ of the scattering material;
hence, that reversal should not have any effect on the scattered electric and magnetic field phasors.
Simultaneous reversal of the signs of $\epsr$ and $\mur$ in case (b)  does not change
$\etar$ but $k/\ko$ does change; hence, that reversal should affect the scattered field
phasors.

\begin{figure}[htb]
 \centering 
     \begin{subfigure}[h]{0.4\textwidth}
\includegraphics[width=0.8\linewidth]{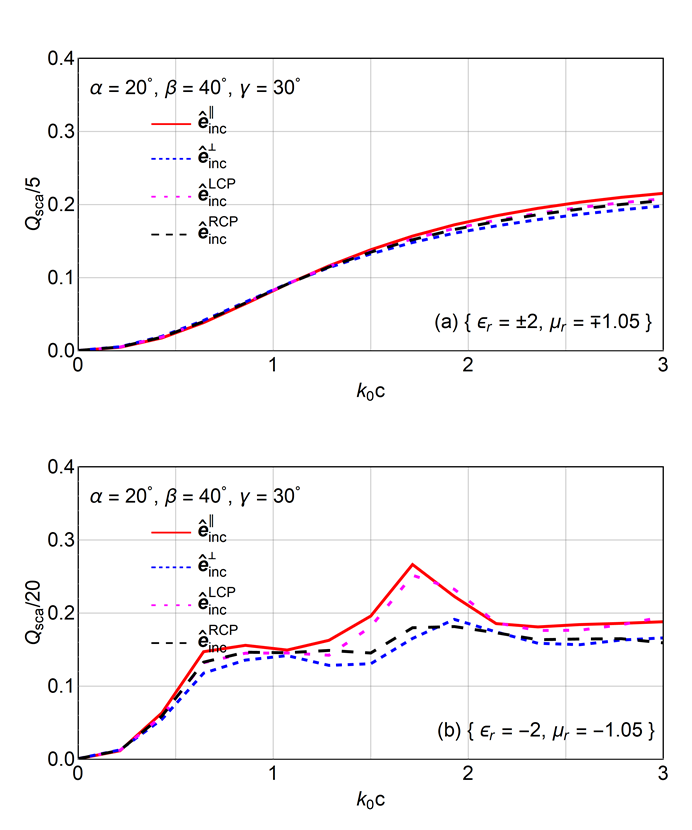}
 \end{subfigure}
\caption{ $\Qscaparl$, $\Qscaperp$, $\QscaLCP$,
and $\QscaRCP$ vs. $\koc\in[0,3]$ for  a biaxially dielectric-magnetic ellipsoid when  $\alpha_x=1.2$, $\alpha_y=1.1$, $\alpha=20^\circ$, $\beta=40^\circ$, and $\gamma=30^\circ$, $a/c=1/2$, and $b/c=2/3$. The angles of incidence are $\theta_{\text{inc}}=45^\circ$ and $\phi_{\text{inc}}=30^\circ$.   
(a) $\epsr=\pm2$ and $\mur=\mp1.05$, (b) $\epsr=-2$, and $\mur=-1.05$.   }
\label{Qsca_general_negative}
\end{figure}

Figure~\ref{Qsca_general_negative}(a) shows plots of  
$\Qscaparl$, $\Qscaperp$, $\QscaLCP$,
and $\QscaRCP$ vs. $\koc$ for the  biaxially dielectric-magnetic ellipsoid of Fig.~\ref{Qsca_general}(d), except that either $\lec \epsr=2,\mur=-1.05\ric$
or $\lec \epsr=-2,\mur=1.05\ric$. As predicted in the previous paragraph, the plots for
$\lec \epsr=2,\mur=-1.05\ric$ and
 $\lec \epsr=-2,\mur=1.05\ric$ are identical, and these plots differ from those for
 $\lec \epsr=2,\mur=1.05\ric$. Furthermore, $\Qsca$ increases with $\koc$ in Fig.~\ref{Qsca_general_negative}(a)  regardless of the polarization state
 of the incident plane wave, but a similar trend is not followed strictly
 in Fig.~\ref{Qsca_general}(d).

Figure~\ref{Qsca_general_negative}(b) shows plots of  $\Qscaparl$, $\Qscaperp$, $\QscaLCP$,
and $\QscaRCP$ vs. $\koc$ for the  biaxially dielectric-magnetic ellipsoid of Fig.~\ref{Qsca_general}(d), except that   $\lec \epsr=-2,\mur=-1.05\ric$. Again,
as predicted   two paragraphs previously, the plots for
$\lec \epsr=-2,\mur=-1.05\ric$ are not the same
as for
 $\lec \epsr=2,\mur=1.05\ric$. Also, regardless of the polarization state
 of the incident plane wave, $\Qsca$ has a complicated dependence on $\koc$ in Fig.~\ref{Qsca_general_negative}(b), but for the most part $\Qsca$  increases
with $\koc$  
in Fig.~\ref{Qsca_general}(d).  Compared to ellipsoids characterized
by either $\lec \epsr>0,\mur>0\ric$ or $\lec\epsr\gtrless0,\mur\lessgtr0\ric$, the 
difference between $\QscaRCP$ and $\QscaLCP$ is much more pronounced for the
ellipsoid characterized by $\lec \epsr<0,\mur<0\ric$.

Suppose that $\vert\epsr\vert$, $\vert\mur\vert$, and $\koc$ are fixed. Then, 
 for a fixed polarization state of the incident plane wave, $\Qsca$ is maximum for
$\lec \epsr<0,\mur<0\ric$, intermediate for  $\lec\epsr\gtrless0,\mur\lessgtr0\ric$, and
minimum for $\lec \epsr>0,\mur>0\ric$, according to Figs.~\ref{Qsca_general}(d)
and \ref{Qsca_general_negative}(a,b).  
Calculations for several other values of
$\vert\epsr\vert$ and $\vert\mur\vert$ are in agreement with this conclusion. 

Finally, in this section we consider the effect of dissipation in the scattering material. The calculations for Fig.~\ref{Qsca_general}
were repeated but with $\epsr=2+i0.1$ and $\mur=1.05+i0.01$; accordingly, $\Qabs\ne0$. 
As dissipation depressed $\Qsca$ slightly but did not otherwise alter the conclusions drawn
from Fig.~\ref{Qsca_general}, plots of $\Qscaparl$, $\Qscaperp$, $\QscaLCP$,
and $\QscaRCP$ vs. $\koc$ are not presented here.

Figure~\ref{Qabs_general} contains plots of $\Qabsparl$, $\Qabsperp$, $\QabsLCP$,
and $\QabsRCP$ vs. $\koc\in[0,3]$. When the principal axes of $\=C$ and $\=U$
coincide, Fig.~\ref{Qabs_general}(a) shows that $\Qabsparl > \Qabsperp$
and $\QabsRCP \simeq \QabsLCP$ for all $\koc$.  
The excess of $\Qabsparl$ over $\Qabsperp$ increases with a rotation
of $\=C$ about the $z$ axis by $\alpha=20^\circ$,
as becomes evident from comparing Figs.~\ref{Qabs_general}(b) with  \ref{Qabs_general}(a), but
$\QabsRCP \simeq \QabsLCP$ even for $\alpha=20^\circ$.
A subsequent rotation by 
$\beta=40^\circ$ about the new $y$ axis increases the excess of $\QscaRCP$ over $\QscaLCP$;
see  Fig.~\ref{Qabs_general}(c).
This excess is not affected by a further rotation by $\gamma=30^\circ$ about the latest $z$ axis in
Fig.~\ref{Qabs_general}(d). an additional conclusion is that
$\QabsRCP-\QabsLCP$ is more sensitive than $\QscaRCP-\QscaLCP$ 
to the orientational mismatch of $\=C$ and $\=U$.

\subsection{Rotation of $\=C$ about a shape principal axis \label{sec3B}}  

\begin{figure}[htb]
 \centering 
     \begin{subfigure}[h]{0.4\textwidth}
\includegraphics[width=0.8\linewidth]{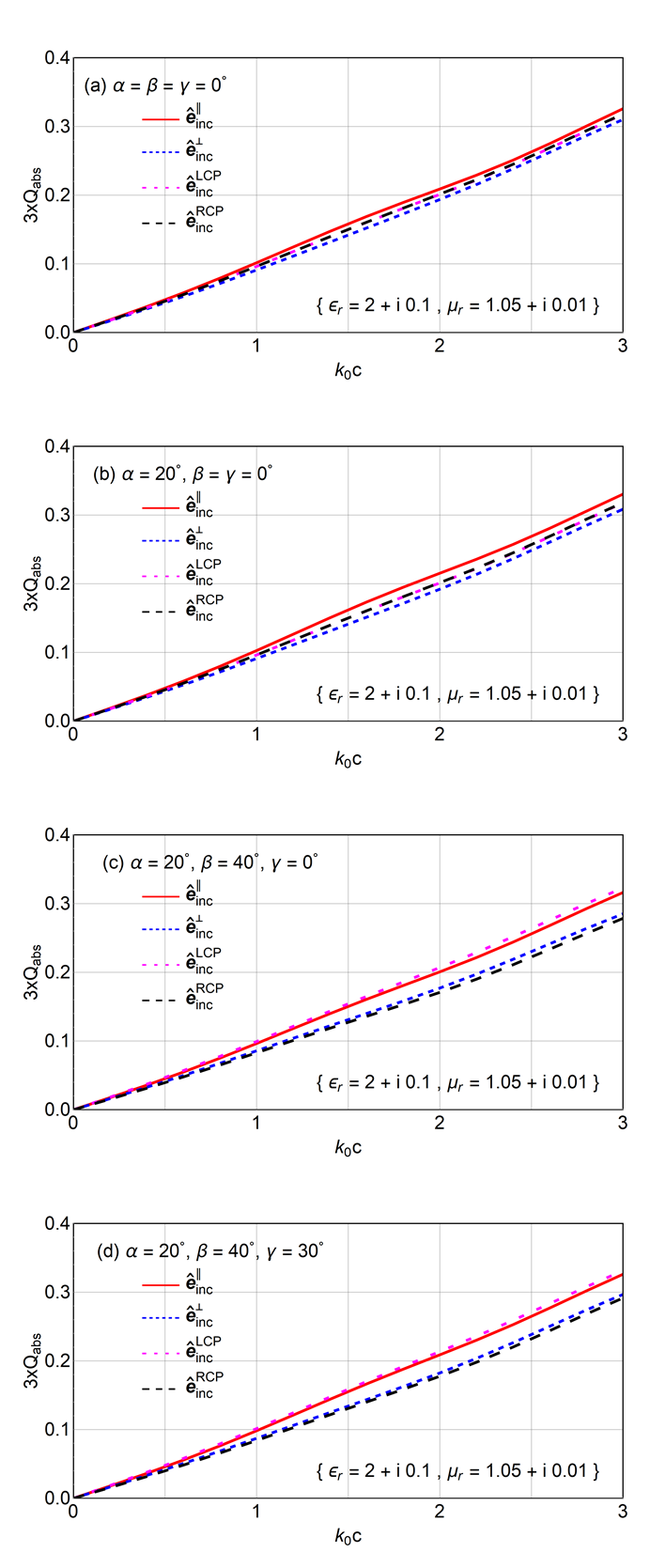}
 \end{subfigure}
\caption{ $\Qabsparl$, $\Qabsperp$, $\QabsLCP$,
and $\QabsRCP$ vs. $\koc\in[0,3]$ for  a biaxially dielectric-magnetic ellipsoid described by  $\epsr=2+i0.1$, $\mur=1.05+i0.01$, $\alpha_x=1.2$, $\alpha_y=1.1$, $a/c=1/2$, and $b/c=2/3$. The angles of incidence are $\theta_{\text{inc}}=45^\circ$ and $\phi_{\text{inc}}=30^\circ$.  
(a) $\alpha=\beta=\gamma=0^\circ$; (b) $\alpha=20^\circ$ and $\beta=\gamma=0^\circ$; (c) $\alpha=20^\circ$, $\beta=40^\circ$, and $\gamma=0^\circ$; and (d) $\alpha=20^\circ$, $\beta=40^\circ$, and $\gamma=30^\circ$.} 
\label{Qabs_general}
\end{figure}

A general rotation of $\=C$, while keeping $\=U$ fixed, requires three consecutive rotations specified
by Eq.~(\ref{S-def}). Each of those rotations has an effect on the scattering and absorption characteristics, as exemplified by Figs.~\ref{Qsca_general} and ~\ref{Qabs_general}. Therefore, we considered
next the variation of $\Qsca$ versus the angle of rotation, but for single rotations; i.e., $\Qsca$ vs. $\alpha$ with $\beta = \gamma= 0^\circ$, and $\Qsca$ vs. $\beta$ with $\alpha = \gamma= 0^\circ$. 
In order to present salient features, we fixed $\epsr=2$ and $\mur=1.05$ for all results presented
in this section; furthermore, as both $\QscaLCP$ and  $\QscaRCP$ are then close to  
  $\left(\Qscaparl+\Qscaperp\right)/2$,
we restricted the incident plane 
wave to be linearly polarized.

\begin{figure}[h]
 \centering 
     \begin{subfigure}[h]{0.5\textwidth}
\includegraphics[width=0.8\linewidth]{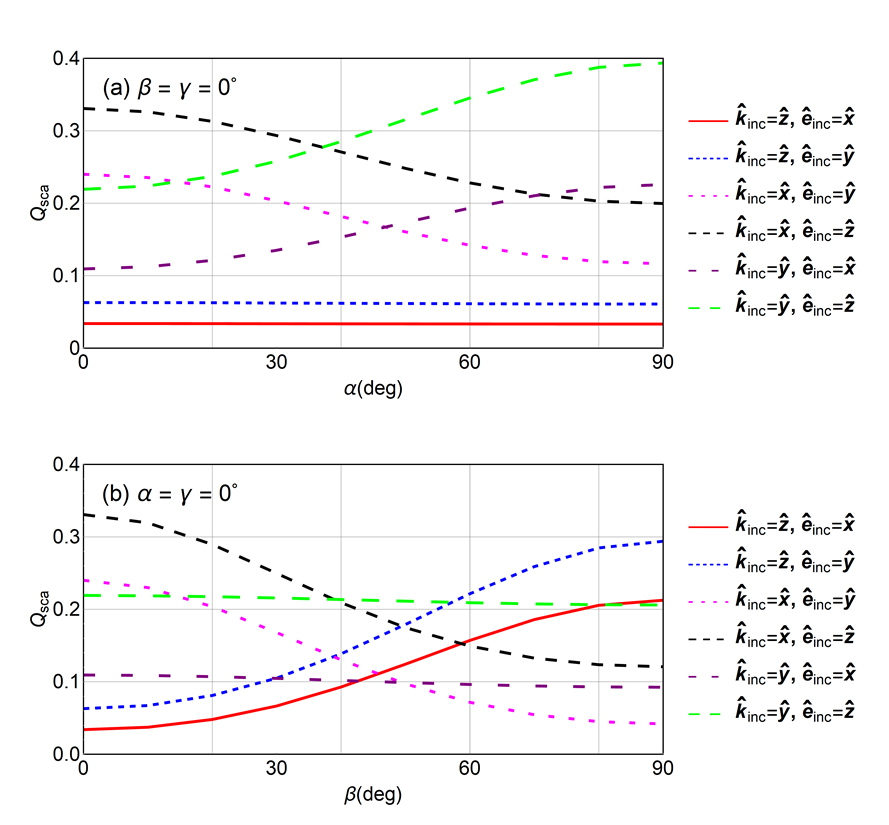}
 \end{subfigure}
\caption{$\Qsca$ vs. $\alpha$ and vs. $\beta$ for  a biaxially dielectric-magnetic ellipsoid when $\epsr=2$, $\mur=1.05$, $\alpha_x=1.2$, $\alpha_y=1.1$, $a/c=1/2$, $b/c=2/3$, and $\koc=3$. 
The constitutive principal axes are  
rotated about either (a) the $z$ axis or (b) the $y$ axis of the laboratory coordinate system.}
\label{Qsca}
\end{figure}

\begin{figure}[h]
 \centering 
     \begin{subfigure}[h]{0.5\textwidth}
\includegraphics[width=0.8\linewidth]{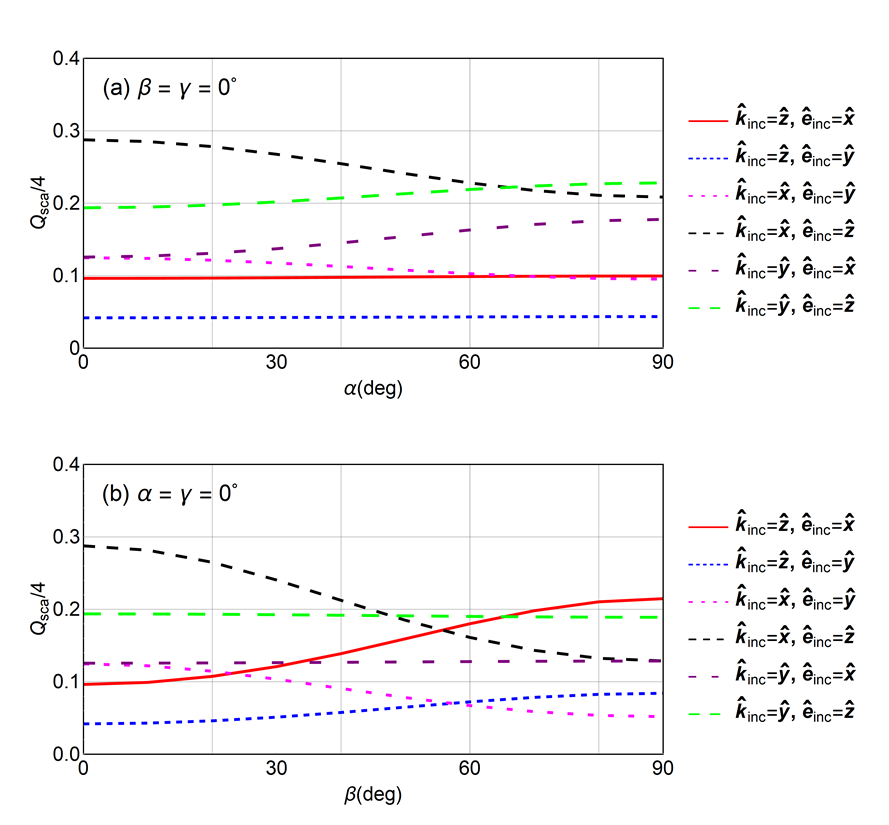}
 \end{subfigure}
\caption{Same as Fig.~\ref{Qsca}, except for a spheroid ($a=c$, and $b=2c/3$) when $\koc=2$.}
\label{Qsca_spheroid}
\end{figure}

\begin{figure}[h]
 \centering 
     \begin{subfigure}[h]{0.4\textwidth}
\includegraphics[width=0.8\linewidth]{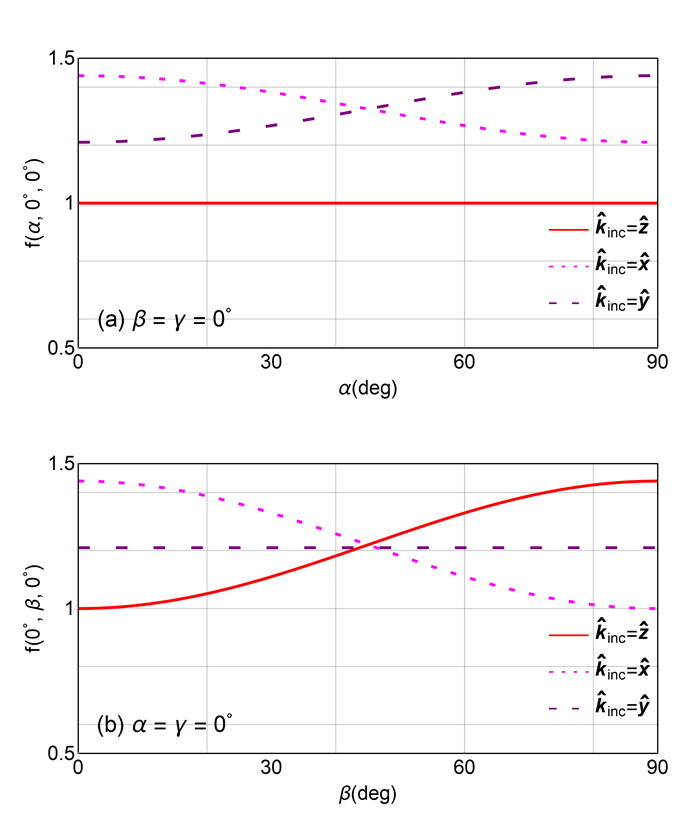}
 \end{subfigure}
\caption{(a) $f(\alpha, 0^\circ, 0^\circ)$ vs. $\alpha$ and (b) $f(0^\circ, \beta, 0^\circ)$ vs. $\beta$,
when $\alpha_x=1.2$, and $\alpha_y=1.1$. }
\label{kbk}
\end{figure}

\begin{figure}[h]
 \centering 
     \begin{subfigure}[h]{0.5\textwidth}
 \includegraphics[width=0.8\linewidth]{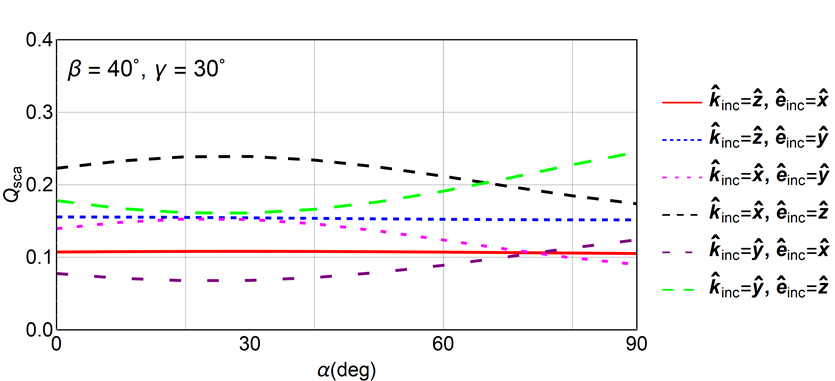}
\includegraphics[width=0.62\linewidth]{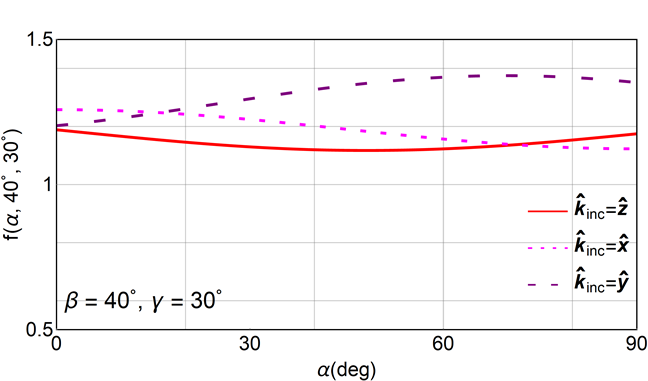}
 \end{subfigure}
\caption{(Top) $\Qsca$ vs. $\alpha$ for $\beta=40^\circ$ and $\gamma=30^\circ$
and (bottom)  $f(\alpha,40^\circ, 30^\circ)$ vs. $\alpha$,
when $\alpha_x=1.2$ and $\alpha_y=1.1$.  }
\label{kbk_compound}
\end{figure}

\begin{figure}[h]
 \centering 
     \begin{subfigure}[h]{0.5\textwidth}
\includegraphics[width=0.8\linewidth]{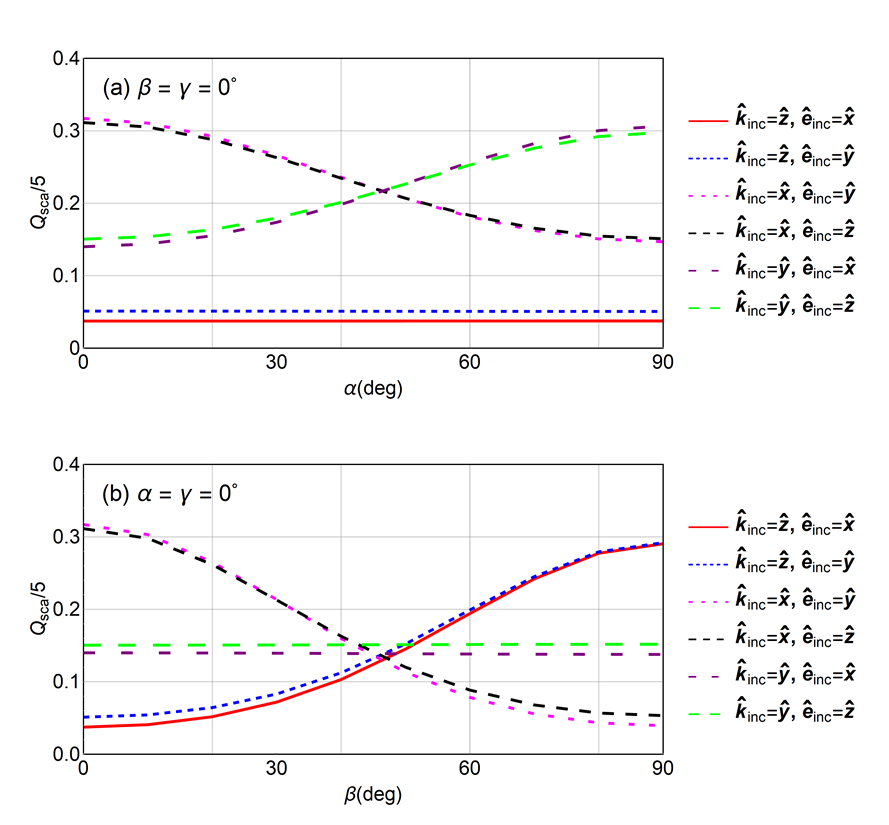}
 \end{subfigure}
\caption{Same as Fig.~\ref{Qsca}, except for a sphere ($a=b=c$).}
\label{Qsca_sphere}
\end{figure}

In Figs.~\ref{Qsca}--\ref{Qsca_sphere}, results are presented for all six canonical configurations of the incident linearly polarized plane wave with respect to the semi-axes of the ellipsoid; i.e., 
$\k_{\text{inc}}\in\lec\ux,\uy,\uz\ric$ and $\e_{\text{inc}}\in\lec\ux,\uy,\uz\ric$ such that $\k_{\text{inc}}\perp\e_{\text{inc}}$. 

Figure~\ref{Qsca}(a) shows plots of 
$\Qsca$ vs. $\alpha$ when $\beta=\gamma=0^\circ$ and Fig.~\ref{Qsca}(b) shows plots of 
$\Qsca$ vs. $\beta$ when $\alpha=\gamma=0^\circ$, 
for a biaxially dielectric-magnetic ellipsoid when $\epsr=2$, $\mur=1.05$, $\alpha_x=1.2$, $\alpha_y=1.1$, $a/c=1/2$, $b/c=2/3$, and $\koc=3$. Thus, the constitutive principal axes are  
rotated about the $z$ axis of the laboratory coordinate system for Fig.~\ref{Qsca}(a) 
 and about   the $y$ axis of the laboratory coordinate system for Fig.~\ref{Qsca}(b).
 
 When $\k_{\text{inc}}$ is parallel to the axis of rotation,   $\Qsca$
has a very weak dependence on the rotation angle, as compared to when $\k_{\text{inc}}$ is perpendicular to the axis of rotation. This characteristic is exemplified  by
\begin{itemize}
\item the plots for $\k_{\text{inc}}=\uz$ in comparison to those
for $\k_{\text{inc}}\in\lec\ux,\uy\ric$  in Fig.~\ref{Qsca}(a) as well as
\item the plots for $\k_{\text{inc}}=\uy$ in comparison to those
for $\k_{\text{inc}}\in\lec\ux,\uz\ric$  in Fig.~\ref{Qsca}(b).
\end{itemize}
This characteristic holds also for spheroids  composed
of the chosen material, as is evident from Fig.~\ref{Qsca_spheroid}.

 Plots of  $\Qsca$ vs. $\alpha$ when $\beta= \gamma=0^\circ$ or $\Qsca$ vs. $\beta$ when $\alpha= \gamma= 0^\circ$ in Figs.~\ref{Qsca} and \ref{Qsca_spheroid} can be predicted from plots of the scalar function
\begin{equation}
f(\alpha, \beta, \gamma)= \k_{\text{inc}} \. \={C}^{-1} \. \k_{\text{inc}}\,,
\end{equation}
 vs. $\alpha$ when $\beta= \gamma= 0^\circ$ or vs. $\beta$ when $\alpha= \gamma= 0^\circ$. These plots are shown in Fig.~\ref{kbk}. 
 When $f(\alpha, 0^\circ, 0^\circ)$ [or $f(0^\circ,\beta, 0^\circ)$] is invariant with respect to
$\alpha$ (or $\beta$),   $\Qsca$ is weakly dependent on $\alpha$ (or $\beta$). Likewise, when $f(\alpha, 0^\circ, 0^\circ)$ [or $f(0^\circ,\beta, 0^\circ)$] 
increases/decreases as  $\alpha$ (or $\beta$)   increases, 
 $\Qsca$ also increases/decreases. Finally, when the value
 of $f(\alpha, 0^\circ, 0^\circ)$ [or $f(0^\circ,\beta, 0^\circ)$] is either a maximum or a minimum, the plot of $\Qsca$ vs. $\alpha$ (or vs. $\beta$) also contains the same extremum. This is true regardless of $\koc$, $\epsr$, $\mur$, $a/c$, and $b/c$, as can be seen on comparing Figs.~\ref{Qsca} and \ref{Qsca_spheroid} with Fig.~\ref{kbk}.

However, this prediction fails for compound rotations; e.g., the variation of $\Qsca$ with respect to $\alpha$ is not predicted by the variation of $f(\alpha, \beta, \gamma)$ with respect to $\alpha$ when
when $\beta \neq 0^\circ$ and/or $\gamma \neq 0^\circ$.  The failure is exemplified  in Fig.~\ref{kbk_compound} by a comparison of
the plots of  $\Qsca$ and  $f(\alpha, 40^\circ, 30^\circ)$ vs. $\alpha$
 for the   biaxially dielectric-magnetic ellipsoid of Fig.~\ref{Qsca}, but when $\beta= 40^\circ$ and $\gamma=30^\circ$.

The foregoing investigation was repeated with the scattering material being dissipative, as in Sec.~3.\ref{sec3A}. The conclusions held for $\Qsca$.  However, 
neither the variation of $\Qabs$ with $\alpha$ could  be predicted from
$f(\alpha, 0^\circ, 0^\circ)$ nor
 the variation of $\Qabs$ with $\beta$ could  be predicted from
 $f(0^\circ,\beta, 0^\circ)$.

The shape effect can also be appreciated by comparisons with a sphere. Figure~\ref{Qsca_sphere}(a) shows plots of $\Qsca$ vs. $\alpha$, and Fig.~\ref{Qsca_sphere}(b) shows plots of $\Qsca$ vs. $\beta$, both when $a=b=c$. For fixed $\k_{\text{inc}}$, $\e_{\text{inc}}$ has a noticeable effect on the curves of $\Qsca$ vs. $\alpha$, as well as on $\Qsca$ vs. $\beta$ when $a \neq b \neq c$ compared to when $a=b=c$. For example, the curves of $\Qsca$ vs. $\alpha$ for
\begin{itemize}
\item
$ \{ \k_{\text{inc}}= \ux, \e_{\text{inc}}=\uy \}$ and 
\item $ \{ \k_{\text{inc}}= \ux, \e_{\text{inc}}=\uz \}$ 
\end{itemize}
are very close to each other when $a=b=c$ [Fig.~\ref{Qsca_sphere}(a)] but not when $a \neq b \neq c$ [Fig.~\ref{Qsca}(a)]. When the same exercise   was conducted with the scattering material being dissipative,   the conclusion was found to hold for both $\Qsca$ and   $\Qabs$.

\section{Concluding Remarks}\label{s3}
We derived expressions of the vector spherical wavefunctions for an orthorhombic dielectric-magnetic material with a permeability dyadic that is a scalar multiple of the permittivity dyadic and with  constitutive principal axes that are arbitrarily oriented in the laboratory coordinate system. These expressions were used in formulating the extended boundary condition method  for scattering by  a three-dimensional object composed of the chosen material.
Numerical results were obtained for plane-wave scattering 
by an ellipsoid  composed of the chosen material, and the effect of the
rotation of the constitutive principal axes relative to the shape principal axes was studied.   

The following conclusions were drawn from the numerical results obtained by us:
\begin{itemize}
\item The difference between $\Qscaparl$ and $\Qscaperp$ is much
larger than the difference between
$\QscaLCP$  and $\QscaRCP$.
The difference between
$\QscaLCP$  and $\QscaRCP$
  increases as the orientational mismatch between the constitutive principal axes 
and the shape principal axes becomes more pronounced and/or the electrical
size of the ellipsoid increases.
 \item The difference between $\Qabsparl$ and $\Qabsperp$ is much
larger than the difference between
$\QabsLCP$  and $\QabsRCP$, if the scattering material is dissipative.
The difference between
$\QabsLCP$  and $\QabsRCP$
  increases as the orientational mismatch between the constitutive principal axes 
and the shape principal axes becomes more pronounced and/or the electrical
size of the ellipsoid increases.
\item When the scattering material is impedance-matched (i.e., $\epsr=\mur$) to the surrounding free space, the differential scattering efficiency does not depend on the polarization state of the incident plane wave, regardless of the orientational mismatch between the constitutive principal axes 
and the shape principal axes. This independence extends to the total scattering, extinction,
and absorption efficiencies as well.
\item The dependence of   $\Qsca$ on the angle of rotation about a shape principal axis can be predicted qualitatively from the variation of the scalar $\k_{\text{inc}} \. \=C \. \k_{\text{inc}}$ with respect to the angle of rotation, regardless
of $\epsr$, $\mur$,  $a/c$, $b/c$, and $\koc$.
When $\k_{\text{inc}}$ is parallel to the axis of rotation, $\Qsca$
does not vary considerably with respect to the angle of rotation,
as compared to when $\k_{\text{inc}}$ is
perpendicular to that axis.
\item When the scattering material is nondissipative,
$\Qsca$ is maximum  when both   $\epsr$ and   $\mur$ are  negative,
intermediate when the two have opposite signs, and minimum
when both are
positive, regardless of the orientational mismatch between the constitutive principal axes 
and the shape principal axes.
\item  Both $\Qsca$  and $\Qabs$
are more sensitive to the polarization state of the incident plane wave for ellipsoids and spheroids than for
spheres.
\end{itemize}

\vspace{0.5cm}
\noindent {\bf Acknowledgments.}  The authors are grateful to an anonymous reviewer for comments that led to the validation of our EBCM program against data obtained using the FDTD method and the FEM.  AL thanks the Charles Godfrey Binder Endowment at Penn State for ongoing support of his research
activities.


\begin{thebibliography}{99}

\bibitem{Post}
E. J. Post,
\textit{Formal Structure of Electromagnetics}
(North-Holland, 1962).

\bibitem{Lakhtakiaan} 
T. G. Mackay and A. Lakhtakia,
\textit{Electromagnetic Anisotropy and Bianisotropy: A Field Guide} 
(World Scientific, 2010).

\bibitem{Chen}
H. C. Chen,
\textit{Theory of Electromagnetic Waves}
(McGraw--Hill, 1983).

\bibitem{Berry}
M. V. Berry,
``The optical singularities of bianisotropic
crystals," Proc. R. Soc. Lond. A \textbf{461}, 2071--2098 (2005).

\bibitem{collin} 
R. E. Collin,
\textit{Field Theory of Guided Waves} 
(IEEE Press, 1991).

\bibitem{FLbook}
M. Faryad and A. Lakhtakia,
\textit{Infinite-Space Dyadic Green Functions in Electromagnetism}
(Morgan \& Claypool, 2018).


\bibitem{Lakhtakiaoriginal} 
A. Lakhtakia and T. G. Mackay,
``Vector spherical wavefunctions for orthorhombic dielectric-magnetic material with gyrotropic-like magnetoelectric properties," J. Opt. (India) \textbf{41}, 201--213 (2012).

\bibitem{Waterman} 
P. C. Waterman,
``Matrix formulation of electromagnetic scattering," Proc. IEEE \textbf{53}, 805--812 (1965).

\bibitem{Lakh-IskBook}
A. Lakhtakia,  ``The Ewald--Oseen extinction theorem
and the extended boundary condition method,'' in \textit{The World of Applied Electromagnetics}, A. Lakhtakia and 
C.~M. Furse, eds. (Springer, 2018), pp.~481--513.

\bibitem{Morse} 
P. M. Morse and H. Feshbach,
\textit{Methods of Theoretical Physics, Vol. II} 
(McGraw--Hill, 1953), pp.~1865--1866.

\bibitem{BY1975}
P. Barber and C. Yeh,
``Scattering of electromagnetic waves by arbitrarily
shaped dielectric bodies," Appl. Opt. \textbf{14}, 2864--2872 (1975).

\bibitem{Lutkepohl}
H. L\"utkepohl, 
\textit{Handbook of Matrices} (Wiley, 1996).


\bibitem{WC}
K. W. Whites and C. Y. Chung,
``Composite uniaxial bianisotropic chiral materials characterization: Comparison
of predicted and measured scattering," {J. Electromagn. Waves Appl.} \textbf{11}, 371--394 (1997).

\bibitem{MW2000}
T. G. Mackay and W. S. Weiglhofer,
``Homogenization of biaxial composite
materials: bianisotropic properties," {J. Opt. A: Pure Appl. Opt.}
\textbf{3}, 45--52 (2000).

\bibitem{MMR}
R. Marqu\'es, F. Medina,   and R. Rafii--El--Idrissi, 
``Role of bianisotropy in negative permeability and left-handed metamaterials,"
{Phys. Rev. B} {\bf 65}, 144440 (2002).

\bibitem{Dionne}
S. N. Sheikholeslami, H. Alaeian, A. L. Koh, and J. A. Dionne,
``A metafluid exhibiting strong optical magnetism,"
{Nano Lett.} {\bf 13}, 4137--4141 (2013).

\bibitem{Plebanski}
J. Pl\'ebanski, ``Electromagnetic waves in gravitational fields," Phys.
Rev. \textbf{118}, 1396--1408 (1960).


\bibitem{Alkhoori1}
H. M. Alkhoori, A. Lakhtakia, J.~K. Breakall, and C.~F. Bohren,
``Plane-wave scattering by an ellipsoid composed
of an orthorhombic dielectric--magnetic material,"
\textit{J. Opt. Soc. Am. A} {\bf 35}, 1549--1559 (2018).

\bibitem{Jafri} 
A. D. U. Jafri and A. Lakhtakia,
``Scattering of an electromagnetic plane wave by a homogeneous sphere made of an orthorhombic dielectric-magnetic material," J. Opt. Soc. Am. A \textbf{31}, 89--100 (2014).

\bibitem{Jafri-err} 
A. D. U. Jafri and A. Lakhtakia,
``Scattering of an electromagnetic plane wave by a homogeneous sphere made of an orthorhombic dielectric-magnetic material: erratum," J. Opt. Soc. Am. A \textbf{31}, 2630 (2014).


\bibitem{Bohren} 
C. F. Bohren and D. R. Huffman,
\textit{Absorption and Scattering of Light by Small Particles} 
(Wiley, 1983).


\bibitem{Aydin}
K. Aydin and A. Hizal, ``On the completeness of the spherical
vector wave functions, " {J. Math. Anal. Appl.} {\bf 117},
428--440 (1986).

\bibitem{Stratton} 
J. A. Stratton,
\textit{Electromagnetic Theory} 
(McGraw--Hill, 1941), pp.~564--565.

\bibitem{Bowman}
J. J. Bowman, T. B. A. Senior, and P. L. E. Uslenghi, eds.,
\textit{Electromagnetic and Acoustic Scattering by Simple Shapes}  
(North-Holland, 1969).

\bibitem{Karam1982}
M. A. Karam and A. K. Fung, ``Vector forward scattering theorem,"
\textit{Radio Sci.} {\bf 17}, 752--756 (1982).

 

\bibitem{Baum2007}
C. E. Baum, ``The forward-scattering theorem applied to the
scattering dyadic,"
\textit{IEEE Transactions on Antennas and Propagation} {\bf 55}, 1488--1494
(2007).




\bibitem{Alkhoori2} 
H. M. Alkhoori, A. Lakhtakia, J. K. Breakall and C. F. Bohren,
``Scattering by a three-dimensional object
composed of the simplest
Lorentz-nonreciprocal medium,"  {J. Opt. Soc. Am. A} {\bf 35},  2026--2034 (2018). 




\bibitem{Jaluria}
Y. Jaluria, 
\textit{Computer Methods for Engineering} (Taylor \& Francis,  1996), Sec.~7.5.3.

\bibitem{GS}
I. S. Gradshteyn and I. M. Ryzhik, \emph{Table of Integrals, Series, and Products} 7th edn.
(Academic Press, 2007).

\bibitem{math} 
E. Kreyszig,
\textit{Advanced Engineering Mathematics} 10th edn.
(Wiley, 2011), Sec. 20.2.

\bibitem{FEKO}
{Altair,
\href{https://altairhyperworks.com/product/FEKO}{Feko Overview} (9 April 2019).}

\bibitem{FDTD_book}
{A. Taflove and S. C. Hagness,
{\it Computational Electrodynamics: The Finite-Difference Time-Domain Method, 3rd ed.}
(Artech House, 2005).}

\bibitem{Monk_book}
{P. Monk,
\textit{Finite Element Methods for Maxwell's Equations} (Oxford University Press, 2003).}

\bibitem{Barberbook} 
P. W. Barber and S. C. Hill,
\textit{Light Scattering by Particles: Computational Methods}  
(World Scientific, 1990).







\end{thebibliography}
\end{document}